\documentclass[%
 aip,
 amsmath,amssymb,
 reprint,%
nofootinbib]{revtex4-2}

\usepackage{graphicx}
\usepackage{dcolumn}
\usepackage{bm}
\usepackage{color, colortbl}
\usepackage{booktabs} 
\usepackage{siunitx}  
\usepackage{tabularx}
\definecolor{Gray}{gray}{0.9}
\usepackage{hyperref}

\begin{document}

\title{Data‑Driven Molecular Dynamics and TEM Analysis of Crystal Growth and Hydrogen Sensing in Pt‑Functionalized Graphene Chemiresistive Sensors}

\author{Akram Ibrahim}
\affiliation{%
Department of Physics, University of Maryland Baltimore County, Baltimore, Maryland 21250, United States
}%

\author{Ahmed M. Hafez}
\affiliation{%
Center for Research and Exploration in Space Science and Technology, NASA Goddard Space Flight Center, Greenbelt, Maryland 20771, United States}%
\affiliation{%
Department of Astronomy, University of Maryland College Park, College Park, Maryland 20742, United States}%

\author{Mahmooda Sultana}
\affiliation{%
Planetary Environments Laboratory, NASA Goddard Space Flight Center, Greenbelt, Maryland 20771, United States}%

\author{Can Ataca}
\affiliation{%
Department of Physics, University of Maryland Baltimore County, Baltimore, Maryland 21250, United States
}%

\email{ataca@umbc.edu}

\date{\today}

\begin{abstract}
Graphene functionalized with catalytic transition metals offers high-performance gas sensing by coupling graphene’s exceptional electronic transport properties with the metal’s catalytic activity, yet the atomistic relationships connecting synthesis parameters, morphological outcomes, and sensor performance remain elusive. We developed an equivariant machine-learning interatomic potential (MLIP) with near-DFT accuracy to perform large-scale molecular dynamics (MD) simulations of Pt crystal growth on graphene and subsequent H$_2$-sensing. MD simulations validated by TEM show that Pt deposition begins with dispersed nuclei coalescing into polycrystalline nanoclusters with predominantly FCC interiors, while both MD and Raman spectroscopy reveal a predominantly non-covalent Pt–graphene interaction that induces moderate local strain and charge transfer yet preserves graphene’s structural integrity. Reactive MD simulations confirm low-barrier H$_2$ dissociative chemisorption exclusively on Pt, with negligible spillover or chemical interaction with pristine graphene. However, H adsorption on Pt attenuates the Pt–graphene interfacial binding, providing an indirect electronic pathway for H$_2$ sensing. Transient and steady‐state kinetic analyses demonstrate that an intermediate Pt loading minimizes the limit of detection; lower loadings facilitate faster response and recovery kinetics and enhance signal transduction, whereas higher loadings increase the doping level of graphene. DFT charge analysis indicates that under-coordinated Pt clusters induce $n$-type doping in graphene, whereas continuous Pt films induce $p$-type doping, with both effects attenuated upon H adsorption. The developed machine‐learned MD framework enables quantum-mechanically accurate modeling of metal crystal growth on graphene, elucidates the underlying H$_2$-sensing mechanism, and correlates several key sensing figures of merit with metal loading and morphology, thereby establishing a predictive pipeline from synthesis conditions to device performance and facilitating \emph{in silico} optimization of chemiresistive gas sensors.
\end{abstract}

\maketitle
Recent years have witnessed rapid advances in gas‐sensing technology, driven by its expanding applications in industrial processes, safety and health monitoring, and space exploration~\cite{acharyya2020functionalization, bulemo2025selectivity, yang2017gas, silva2024gas}. Among gaseous species, hydrogen is particularly important because of its extensive industrial applications in fuel cells, petrochemical refining, and ammonia synthesis \cite{singla2021hydrogen, manna2021opportunities}, as well as its key role in planetary atmospheres \cite{mikal2023hubble}. However, its inherent flammability, combined with its colorless, odorless, and buoyant nature, presents serious safety risks \cite{buttner2011overview}. Among sensor modalities, chemiresistive devices—which transduce surface chemical reactions into resistance changes—are especially promising, offering high sensitivity, low-cost fabrication, and portability \cite{chiu2013towards, koo2020chemiresistive, koo2019metal}. 

Recently, there have been significant advances in carbon-based chemical sensors, particularly for carbon nanotube and graphene-based chemiresistors \cite{schroeder2018carbon, wang2016review, toda2015recent}. Graphene’s high surface-to-volume ratio, zero bandgap, and exceptional carrier mobility exceeding \(10^4\ \mathrm{cm}^2/\mathrm{V.s}\) with low electrical noise make it an ideal platform for chemiresistive gas sensing, as even trace adsorbates produce measurable shifts in Fermi level and conductance \cite{recum2024graphene, bolotin2008ultrahigh, basu2012recent}. However, its perfect \(sp^2\) honeycomb lattice is chemically inert, offering few adsorption sites and thus limited sensitivity and selectivity \cite{gao2016heat, ye2022selective}. To overcome this, surface functionalization is essential—but must preserve graphene’s $\pi$-conjugation and minimal modification depth to retain its electronic advantages \cite{BOSCHNAVARRO201752, georgakilas2016noncovalent, acharyya2020functionalization}. Even low‐level covalent bonding rapidly degrades mobility, so non-covalent approaches that leave the graphene backbone intact are generally favored \cite{marsden2015effect}.

A variety of functionalization strategies---including polymers, metal oxides, and metallic nanoparticles---have been developed to introduce active sites and tailor binding on graphene~\cite{leve2022synergistic, liu2024atomic, acharyya2020functionalization}. In particular, catalytic transition-metal nanostructures (Pt, Pd, Ag) furnish dissociative sites for target gases and, in concert with graphene's conductivity, enable rapid electron transfer~\cite{singhal2017noble, meng2015graphene, acharyya2020functionalization}. Although many studies describe these nanoparticles as purely non-covalently attached, the precise bonding mechanism often remains unverified~\cite{georgakilas2016noncovalent}. Among transition metals, platinum stands out: it efficiently dissociates H$_2$ at moderate temperatures to form surface hydrides (PtH$_x$), which alters the Pt work function and thus Pt--graphene charge transfer and sensor resistivity~\cite{article, kim2014selective}. These hydrogen atoms can either remain on the Pt clusters or, depending on cluster size and metal-substrate interactions, spill over onto the graphene lattice~\cite{olsen1999atomic, christmann1976adsorption, sihag2019dft}. When spillover occurs, adsorbed hydrogen directly modifies graphene's carrier density and scattering characteristics. Even without significant spillover, however, H$_2$ adsorption on the metal surface can still modulate the metal--graphene interface, transmitting a measurable electronic response to graphene, as revealed in this study.

A core challenge in advancing chemiresistive gas sensors lies in accurately modeling the reactive sensing kinetics between the device and the gas. An additional challenge with metal-functionalized graphene sensors is that the device’s underlying crystal structure is unknown. Pt structures on graphene reflect a balance between nondirectional Pt–Pt cohesion and Pt–graphene interfacial binding~\cite{ruffino2017review}. Consequently, Pt often forms three-dimensional (3D) nanoclusters to maximize coordination~\cite{ferbel2024platinum, kim2021tailored}, while strong covalent Pt–C interactions can produce atomically thin metal domains~\cite{robertson2019atomic, abdelhafiz2015layer}. This adds an extra layer of complexity, rendering it a two-fold problem. On one hand, understanding nanoscale Pt nucleation and domain coalescence, as well as how metal loading influences morphology and graphene binding, is essential for optimizing sensor crystal structures and graphene doping. On the other hand, understanding device–gas interactions at initial exposure, during steady state, and throughout gas purge enables the extraction of key performance metrics—sensitivity, limit of detection, response time, and recovery time—and elucidation of the sensing mechanism.

Despite extensive experimental and computational efforts, atomistic understanding of how Pt nanostructures grow on graphene and modulate H$_2$-sensing kinetics under realistic conditions remains limited. This gap stems from the restricted spatiotemporal resolution of \emph{in situ} probes (slow scan rates in scanning tunneling microscopy, ensemble averaging in X-ray absorption, insufficient atomic resolution in infrared spectroscopy, etc.) and from the prohibitive time and length scales of \emph{ab initio} methods such as density-functional theory (DFT)~\cite{owen2024atomistic}. Consequently, previous experimental studies could only correlate metal loading and as-synthesized Pt crystal geometries with sensor response, rather than elucidate definitive atom-level sensing mechanisms~\cite{chu2011effect, kim2014selective, phan2019high, ferbel2024platinum}. Likewise, computational work has been confined to classical adsorption–desorption frameworks or to simplified DFT treatments of a few adatoms and highly symmetric nanoclusters on graphene interacting with gas molecules~\cite{ghosh2017modeling, singla2021effect, kishnani2021palladium, wang2019characterization, felix2024investigating}. 

In this work, we move beyond pre-assumed crystal structures and the size limitations of DFT to present a comprehensive atomistic dynamics study at experimentally relevant length- and time-scales. To this end, we develop and apply a machine-learned interatomic potential (MLIP) based on local equivariant deep neural networks, which construct a many-body potential through iterated tensor products of learned equivariant representations of the atomic environments~\cite{musaelian2023learning, batzner20223, geiger2022e3nn}. MLIPs have recently gained prominence in computational materials science for their near-DFT accuracy, nearly linear scaling with system size, and excellent transferability across diverse chemical environments, with demonstrated applications in modeling material synthesis and gas–surface interactions~\cite{behler2021machine, friederich2021machine, ibrahim2024modeling, chen2024diffusion, vandermause2022active, owen2024atomistic, shaidu2024entropic}. This approach ensures both high fidelity and superior scalability when simulating complex Pt–graphene–hydrogen systems. In parallel with our computational modeling efforts, we have conducted transmission electron microscopy (TEM) and Raman spectroscopy characterizations to examine the morphological characteristics of Pt nanostructures and their interactions with graphene at different loading levels.  

Overall, our machine-learned large-scale MD simulations reveal that Pt loading and deposition method govern morphology, with rapid deposition yielding dispersed nucleation centers that coalesce into polycrystalline 3D nanoclusters with predominantly FCC interiors and other occasional metastable phases—consistent with TEM observations—and layer-by-layer deposition stabilizing atomically thin Pt films. MD and Raman spectroscopy both reveal that the Pt–graphene interaction is predominantly non-covalent, imparting moderate local strain and charge transfer while preserving graphene’s structural integrity. We observe that H$_2$ dissociatively chemisorbs on Pt with negligible spillover onto graphene, yet H adsorption on Pt weakens Pt–graphene binding and thus transduces the sensing signal at the interface. Kinetic analysis reveals that an intermediate Pt loading optimally minimizes the limit of detection; lower loadings accelerate both response and recovery times and enhance signal transduction, whereas higher loadings increase the doping level of graphene. Charge-transfer calculations further demonstrate that under-coordinated Pt clusters induce $n$-type doping in graphene, whereas extended Pt slabs yield $p$-type behavior, and that H adsorption on Pt attenuates graphene doping in both cases. Our findings unveil the atomistic sensing mechanism in Pt–graphene gas sensors and establish precise design rules linking device-scale properties—sensitivity, response, and recovery times—to underlying crystal morphology and chemical synthesis conditions. This framework paves the way for physics-based, machine-learned \emph{in silico} studies of chemiresistive gas sensors—from material synthesis to gas-sensing operation—with quantum-mechanical fidelity at experimentally relevant length and time scales.

\section{Results and Discussion}

\begin{figure*}
    \centering
    \includegraphics[width=0.94\textwidth]{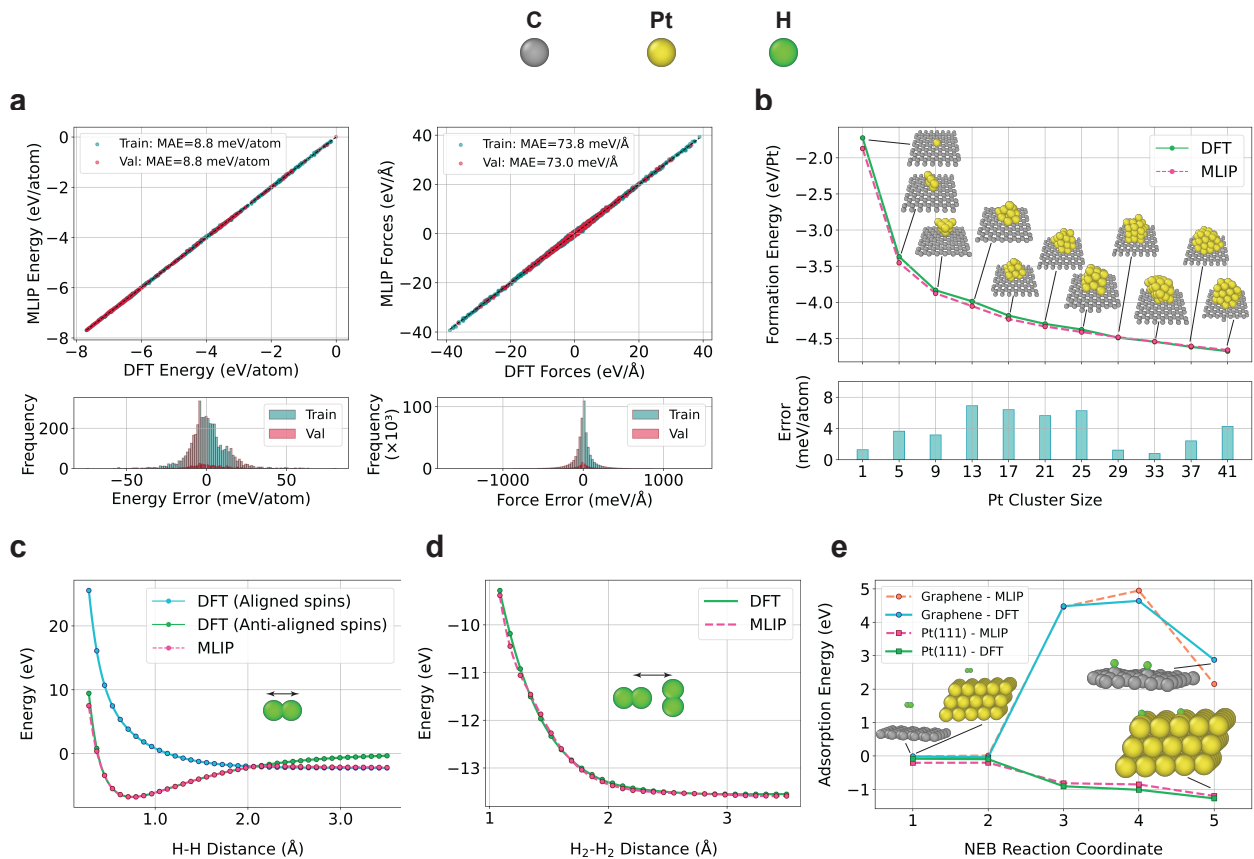}
    \caption{\textbf{Evaluating the Machine-Learned Potential.} \textbf{a} Performance of the MLIP on the training (90\;\%) and validation (10\;\%) datasets for predicting DFT formation energies and atomic forces. \textbf{b} Comparison of MLIP versus DFT for predicting the formation energies (per Pt atom), along with the associated energy errors (normalized by the total number of C and Pt atoms), for Pt clusters of various sizes (optimized on graphene using the minima hopping algorithm). \textbf{c} PESs (total energies) of MLIP versus DFT for H$_2$ molecule with aligned and anti-aligned spins. \textbf{d} MLIP versus DFT PES (total energies) for two perpendicularly oriented H$_2$ molecules. \textbf{e} Adsorption energies of MLIP versus DFT for NEB images along the minimum energy path identified by MLIP, for H$_2$ dissociative adsorption on both graphene and Pt(111) surfaces.}
    \label{fig:fig1}
\end{figure*}

\textbf{Benchmarking the MLIP for modeling Pt / graphene structures and reactivity to H / H$_2$.} We initially demonstrate the generalization capabilities of our MLIP by benchmarking it against DFT calculations. Figure \ref{fig:fig1} outlines multiple aspects of this benchmarking process. Figure \ref{fig:fig1}a illustrates the energies and forces predicted by MLIP for the training structures (90\;\%) and the unseen validation structures (10\;\%) along with the associated error distributions. The MLIP achieves an impressive accuracy with an energy mean absolute error (MAE) $< 9$ meV/atom and forces MAE $< 75$ meV/\AA\; for both training and validation sets. It should be noted that the data set includes a diverse array of configurations, such as Pt-layered and nanocluster structures on graphene, isolated Pt clusters, and both physisorbed H$_2$ and chemisorbed H on various Pt/graphene configurations. This diversity is evidenced by the wide range of atomic forces and formation energies illustrated in Figure \ref{fig:fig1}a.

To further validate the predictive capabilities of the MLIP in optimizing Pt/graphene structures, we utilized the MLIP to conduct minima hopping (MH) optimizations. MH is a global crystal structure optimization method that utilizes a stochastic walker to traverse the potential energy surface (PES), facilitating transitions from one local minimum to another and steering the search toward more energetically favorable structures \cite{krummenacher2024performing}. Figure \ref{fig:fig1}b displays the MH-predicted configurations of Pt$_N$ clusters on pristine graphene for $N = 1,5,9,\dots,41$ ($\Delta N = 4$), along with the corresponding formation energies as calculated by the MLIP and compared with DFT static calculations. The formation energy per Pt atom is defined as:
\begin{equation}
E^{\text{form}}_{\text{Pt}} = \left(E^{\text{total}}_{\text{Gr+Pt}} - E^{\text{total}}_{\text{Gr}} - N_{\text{Pt}} \times E_{\text{Pt}}^{\text{vac}}\right) / N_{\text{Pt}}
\label{eq:formation_energy}
\end{equation}
where $E^{\text{total}}_{\text{Gr+Pt}}$ represents the total energy of graphene with the Pt cluster adsorbed onto it, $E^{\text{total}}_{\text{Gr}}$ denotes the total energy of the relaxed bare graphene without any Pt atoms, $N_{\text{Pt}}$ is the number of Pt atoms in the cluster, and $E_{\text{Pt}}$ refers to the energy of an isolated Pt atom in vacuum. The MLIP demonstrates high accuracy in predicting the energies of Pt clusters on graphene across various sizes, with errors $< 8$ meV/atom, aligning within the established training and validation error bounds detailed in Figure \ref{fig:fig1}a.

Figure \ref{fig:fig1}c underscores the MLIP's accuracy in predicting hydrogen energetics by illustrating the PES for a single hydrogen molecule in two distinct spin states: aligned and anti-aligned. For H-H bond lengths exceeding $\approx 2.1$ \AA, the MLIP predictions closely align with the PES of aligned spins, accurately capturing the chemical dissociative limit where the hydrogen atoms are non-interacting, and no spin flipping occurs. Below this threshold, where the two PESs intersect and the H-H bond formation begins, the MLIP's predictions transition to match the anti-aligned spins PES, effectively capturing the onset of bond formation between hydrogen atoms in accordance with the Pauli exclusion principle. Furthermore, Figure \ref{fig:fig1}d displays the PES of two perpendicular hydrogen molecules, relaxed with respect to their centers of mass, as a function of the distance between these centers, demonstrating the MLIP's high accuracy in modeling the interactions between H$_2$ molecules in the gaseous state.

Finally, to evaluate the MLIP's capability to model the kinetics of hydrogen on both the Pt and graphene surfaces, we employed the nudged elastic band (NEB) method with the climbing image technique utilizing the MLIP (see Figure \ref{fig:fig1}e) \cite{henkelman2000climbing}. The initial state featured an H$_2$ molecule positioned parallel to the surface at a distance exceeding \(3.5 \, \text{\AA}\) above the graphene and Pt(111) surfaces. The final state comprised two hydrogen atoms, each adsorbed atop two surface carbon or platinum atoms, indicative of a dissociative chemisorption state. The energies of the images along the minimum energy path connecting the initial and final states, as computed by the MLIP, were subsequently validated through static DFT calculations. The adsorption energy of hydrogen along the NEB reaction coordinate, depicted in Figure \ref{fig:fig1}e, is defined as:
\begin{equation}
E^{\text{ads}}_{\text{H-Gr/H-Pt}} = E^{\text{total}}_{(\text{Gr/Pt})+\text{H}} - E^{\text{total}}_{\text{Gr/Pt}} - \frac{N_{\text{H}}}{2} \times E_{\text{H}_2}^{\text{vac}}
\label{eq:adsorption_energy}
\end{equation}
where \(E^{\text{total}}_{(\text{Gr/Pt})+\text{H}}\) is the total energy of the system comprising the graphene or Pt substrate with adsorbed hydrogen atoms, \(E^{\text{total}}_{\text{Gr/Pt}}\) is the total energy of the relaxed bare graphene or platinum substrate without adsorbed hydrogen, \(N_{\text{H}}\) is the number of hydrogen atoms, and \(E_{\text{H}_2}^{\text{vac}}\) represents the total energy of an isolated hydrogen molecule in vacuum. 

Figure \ref{fig:fig1}e confirms that the MLIP successfully reproduces the exothermic approximately barrier-less dissociative chemisorption of H$_2$ on the Pt(111) surface~\cite{luntz1990molecular, arboleda2007potential}. Furthermore, the MLIP accurately captures both the endothermic dissociative adsorption of H$_2$ on pristine graphene and the associated high-energy barrier ($\approx 4.6$~eV for the depicted configuration). For reference, the experimental H$_2$ dissociation energy in vacuum is $\approx4.5$~eV~\cite{herzberg1961dissociation}, whereas DFT-calculated barriers on graphene range from $\approx3.3$ to $\approx4.7\,$eV depending on the configuration~\cite{miura2003first}.
\\

\begin{figure*}
    \centering
    \includegraphics[width=1.0\textwidth]{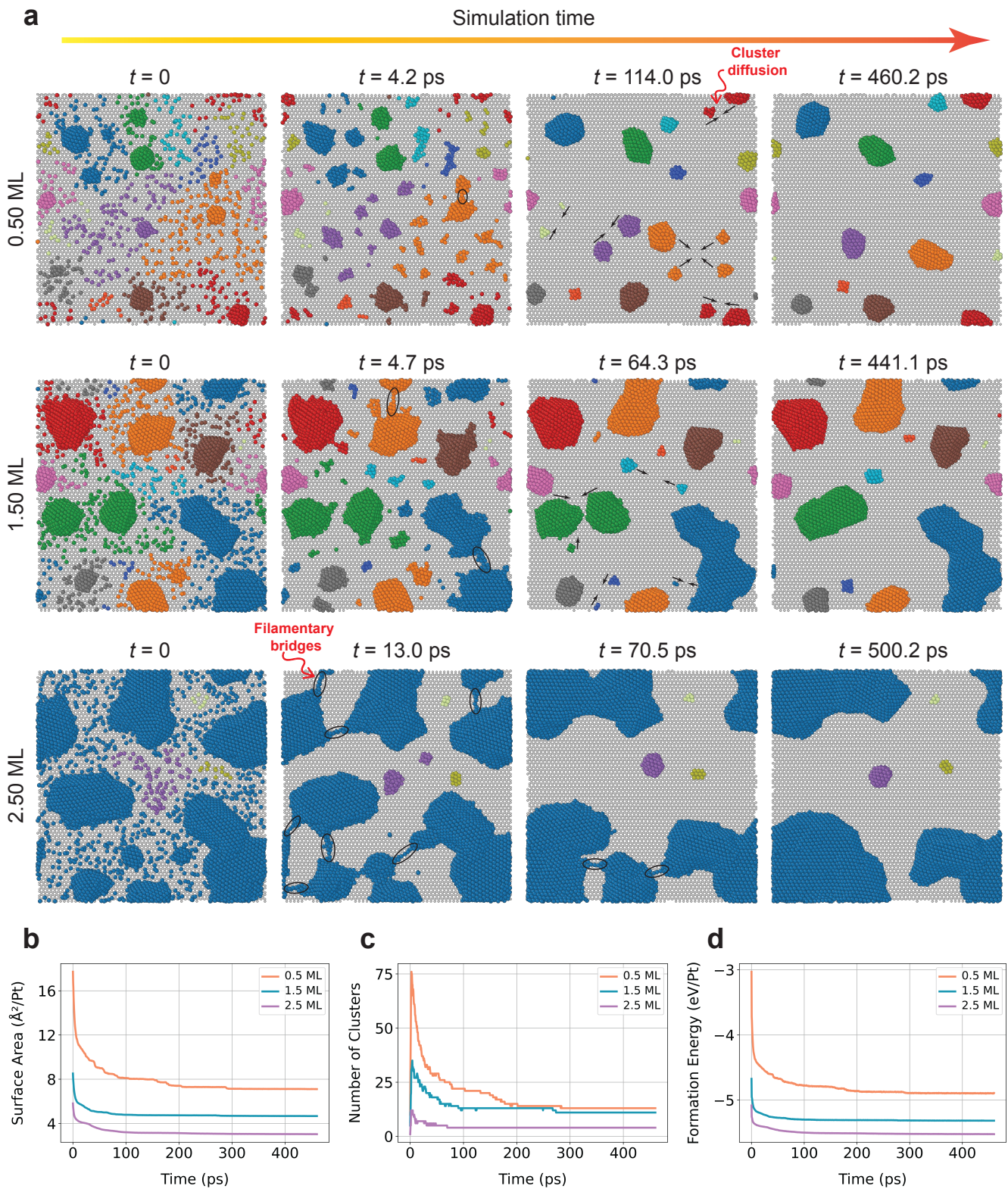}
    \caption{\textbf{Nucleation and Structural Evolution of Pt Nanostructures on Graphene.} \textbf{a} Time evolution of Pt atoms on graphene, shown for 3 Pt loadings ($0.50$ ML, $1.50$ ML, and $2.50$ ML). The initial frame ($t=0$) depicts the baseline configuration, based on the lower‐loading optimized structure and augmented by the deposition of a 0.25~ML batch of Pt atoms. Pt atoms are color-coded to correspond to the final cluster they inhabit. A cutoff distance of $6.5$ \AA~is employed for clustering. \textbf{b, c, d} Trajectories of Pt surface area (per Pt atom), number of unique clusters, and formation energy (per Pt atom) over time.}
    \label{fig:fig2}
\end{figure*}

\textbf{Nucleation and Growth Dynamics of Pt on Graphene.} Under typical physical-vapor deposition (PVD) conditions, Pt atoms are ejected from a solid source and adsorb onto graphene to form primary nuclei. Which crystal morphologies do these nascent nuclei eventually adopt, and how do they evolve over time? To answer this, we employ our MLIP in MD annealing simulations at $300\,$K to provide an atomistic perspective on Pt nucleation and structural evolution during PVD. Simulations are conducted at $300\,$K to emulate the ambient‐temperature condensation on the substrate characteristic of PVD, in contrast to the elevated temperatures used in chemical‐vapor deposition (CVD)~\cite{ALIOFKHAZRAEI201485}. We deposit Pt in sequential increments of $0.25$ monolayer (ML), where $1\,$ML corresponds to a close‐packed Pt overlayer on graphene with a nearest‐neighbor spacing of $2.8\,$\AA{}, consistent with the bulk interatomic spacing (see Figure~S1). In our $60\times35\sqrt{3}$ graphene supercell ($\approx14.8\,$nm$\times15.0\,$nm), $0.25\,$ML corresponds to 812 Pt atoms. We tested several deposition batch sizes and found that $0.25\,$ML provides sufficient mobility for atoms to reach their ground states before the next deposition.

We performed MD simulations of Pt crystal growth across loadings from $0.25\,$ML to $2.50\,$ML. Figure~\ref{fig:fig2}a shows the temporal evolution of depositions at three representative loadings—$0.50\,$ML, $1.50\,$ML, and $2.50\,$ML—on pristine graphene. The different colors assigned to Pt atoms in Figure~\ref{fig:fig2}a denote the unique clusters they form upon reaching equilibrium. For each loading, the $t=0$ frame is generated by adding $0.25\,$ML of Pt to the equilibrated configuration from the previous loading. Simulations were terminated when the total energy plateaued and the morphology remained unchanged for at least $100\,$ps, ensuring that all atoms have reached their ground‐state positions before the next deposition. 

Figure~\ref{fig:fig2}a elucidates the nucleation behavior of Pt atoms on graphene, highlighting distinct pathways for the incorporation of newly deposited atoms. For instance, comparing the initial frame at $t=0$ ps and the subsequent frame at $t=4.2$ ps for the $0.50$ ML loading, one can observe two principal behaviors: some atoms aggregate to form nascent nucleating clusters—for instance, those represented by orange and violet—while others, deposited near pre-existing clusters, migrate toward and coalesce with them, enlarging their sizes, as evidenced by the green and grey clusters. These behaviors also occur concurrently across different clusters: an existing cluster may attract only a fraction of the newly deposited atoms situated in its vicinity, while the remainder agglomerate into new separate clusters. 

Two distinct cluster coalescence mechanisms are evident in Figure~\ref{fig:fig2}a. First, in the second time frame across all Pt loadings, we observe the formation of extended Pt filaments (highlighted by black ovals in Figure~\ref{fig:fig2}a). These filaments act as bridges connecting the previously-formed Pt domains, and their occurrence increases with higher Pt loading. From the third time frames, it becomes apparent that these filaments significantly facilitate the coalescence of the connected domains by providing a direct conduit for their agglomeration. Another mechanism of cluster coalescence is observed in the third time frames for the $0.50$ ML and $1.50$ ML loadings, where small to medium-sized clusters diffuse over the graphene substrate and aggregate to form larger clusters (indicated by black arrows), which are more stable within their surrounding environments over relatively long MD simulation times.

\begin{figure*}
    \centering
    \includegraphics[width=1.0\textwidth]{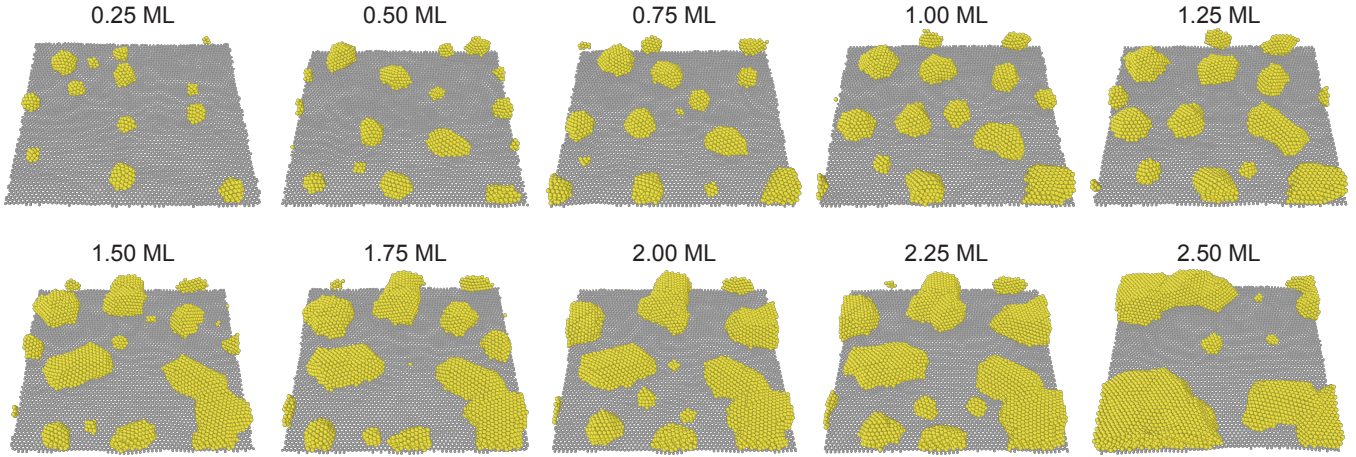}
    \caption{\textbf{Optimized Pt Nanostructures on Graphene.} Pt nanostructures at various loadings were optimized on a 14.8~nm~$\times$~15.0~nm graphene sheet via MD annealing at 300~K with a 2~fs timestep, emulating physical vapor deposition.}
    \label{fig:fig3}
\end{figure*} 

Figures \ref{fig:fig2}b, \ref{fig:fig2}c, and \ref{fig:fig2}d elucidate the temporal progression of several key properties during Pt crystal growth on graphene: the surface area normalized by the number of Pt atoms, the count of Pt clusters (using Ovito’s cluster analysis modifier with a cutoff distance of $6.5$ \AA), and the formation energy per Pt atom [Eq. \eqref{eq:formation_energy}], respectively \cite{stukowski2009visualization}. These figures reveal a gradual decrease in these three properties over time. This trend suggests that Pt atoms, initially dispersed among numerous small clusters, progressively consolidate into fewer, larger clusters. This behavior is underpinned by a fundamental thermodynamic principle: smaller clusters, with their higher surface-to-volume ratios, exhibit elevated surface energies per atom, thereby driving the system to lower its overall free energy through cluster coalescence—a process analogous to classical coarsening \cite{liu2012metals, liu2015growth}.
\\

\textbf{Pt Nanostructure Morphology on Graphene.} Figure~\ref{fig:fig3} presents the optimized structures obtained from our MLIP-driven MD simulations across ten different Pt loadings. All structures exhibit a Volmer–Weber growth mode, where Pt atoms ultimately assemble into 3D nanoclusters dispersed on the graphene surface~\cite{yamakawa2008phase}. Figure~S2 presents quantitative morphological metrics for each Pt loading, including the projected area, mean domain thickness, aspect ratio, minimum and maximum cluster thicknesses, bulk-to-surface atom fractions, and a breakdown of surface sites into terrace and lower-coordination atoms. 

The projected area ($A_{\mathrm{proj}}$) represents the footprint of Pt clusters on graphene (calculated by projecting Pt atomic positions onto the $xy$-plane, identifying clusters via the DBSCAN algorithm \cite{ester1996density}, and applying an alpha-shape algorithm \cite{1056714} to trace each cluster’s boundary and compute its projected area). Figure~S2a illustrates that, despite the increased planar coverage of Pt at higher loadings, the projected area fraction (relative to the graphene area) remains significantly below unity even at $2.50$~ML loading. This indicates incomplete in-plane coverage of the graphene substrate and the persistence of a vertical stacking of Pt atoms. Moreover, Figure~S2a reveals that, as the Pt loading increases, $A_{\mathrm{proj}}$ expands nearly linearly from approximately $10$ to $86$~nm$^2$, corresponding to a threefold increase in the horizontal characteristic length scale (defined as $\sqrt{A_{\mathrm{proj}}}$), from about $3$ to $9$~nm across the $0.25$–$2.50$~ML loading range. This expansion is accompanied by a comparatively smaller, twofold increase in the mean thickness ($\bar t$) of the Pt domains, which rises from approximately $0.5$ to $1$~nm over the same loading range as shown in Figure~S2b. $\bar{t}$ serves as a representative metric for vertical growth, while the distribution of individual cluster thicknesses is shown in Figure S2d. $\bar t$ is calculated as \(\bar{t} = \frac{1}{N_{\mathrm{Pt}}} \sum_{k=1}^{N_{\mathrm{clusters}}} \sum_{i=1}^{N_{\mathrm{Pt},k}} (z_{i,k} - z_{\min,k})\), where \(z_{i,k} - z_{\min,k}\) is the height of atom \(i\) in cluster \(k\) above the lowest atom in that cluster. This per-cluster referencing provides a robust thickness measure, accounting for graphene local corrugations.

Figure~S2c shows the vertical-to-horizontal aspect ratio, a dimensionless quantity defined as $\bar{t} / \sqrt{A_{\mathrm{proj}}}$. This metric serves as an indicator of the balance between vertical and lateral growth. As the Pt loading increases, the aspect ratio decreases sharply up to the $1.00$~ML loading, followed by a more gradual decline up to $2.25$~ML, and then exhibits a pronounced increase at $2.50$~ML. This trend suggests that horizontal growth predominates at lower loadings and continues to dominate at intermediate loadings—albeit at a reduced rate—while at $2.50$~ML, vertical growth becomes more pronounced, as evidenced by the increased thickness in Figure~S2b, indicating the formation of thicker, bulk-like domains. Observations from Figure~\ref{fig:fig2} reveal that newly formed Pt nucleations eventually migrate and adhere to the peripheries of pre-existing clusters. This edge coalescence behavior contributes less to increasing the vertical thickness of Pt domains—resulting in a slowly increasing growth profile in the out-of-plane direction—but contributes more significantly to lateral expansion. Consequently, the vertical-to-horizontal aspect ratio decreases with increasing loading, as shown in Figure~S2c. Figure~\ref{fig:fig3} substantiates these findings by revealing that, while clusters exhibit near-spherical geometries at low Pt loadings, they adopt more flattened morphologies with reduced aspect ratios at higher loadings—also consistent with the morphologies observed in the TEM analysis (Figures~\ref{fig:fig5}a and~\ref{fig:fig5}b). 

From Figure S2e, it is also observed that the number of clusters remains relatively constant (between $11$ and $13$) for coverages ranging from $0.25$ ML to $1.50$ ML, but then declines more rapidly, reaching only $4$ at $2.50$ ML. This trend corroborates that smaller loadings are characterized by a greater number of smaller islands, while increasing loading promotes lateral expansion and enhances the likelihood of edge coalescence into fewer, larger clusters, due to reduced inter-cluster distances. At certain critical loadings, the inter-cluster distances may decrease sufficiently to facilitate significant cluster coalescence, as observed at the Pt loading of $2.50$~ML, as illustrated in Figures~\ref{fig:fig2}a and~\ref{fig:fig3}. Figure~S2f further supports this observation by presenting the sizes of the smallest and largest Pt clusters at each loading (normalized by the total number of Pt atoms). While the minimum cluster size remains nearly constant across the examined range, the largest cluster size shows two notable increases at $1.50$~ML and $2.50$~ML. The latter approaches a value close to unity, indicating a pronounced collective cluster coalescence at the $2.50$ ML loading. 

Lastly, Figure~S2g illustrates that increasing Pt loading elevates the bulk‐to‐surface atom ratio, as quantified by the coordination number (CN) (bulk atoms: $\mathrm{CN}=12$, surface atoms: $\mathrm{CN}<11$).  Concurrently, Figure~S2h reveals that the fraction of terrace sites ($8 \le \mathrm{CN} \le 10$) grows at the expense of lower‐coordination sites such as edge and corner sites ($\mathrm{CN}<8$).  These trends collectively signify a progressive transition toward flatter, facet‐dominated Pt clusters with higher loading, a conclusion that is also visually corroborated in Figure~\ref{fig:fig3}. 
\\

\begin{figure*}
    \centering
    \includegraphics[width=1.0\textwidth]{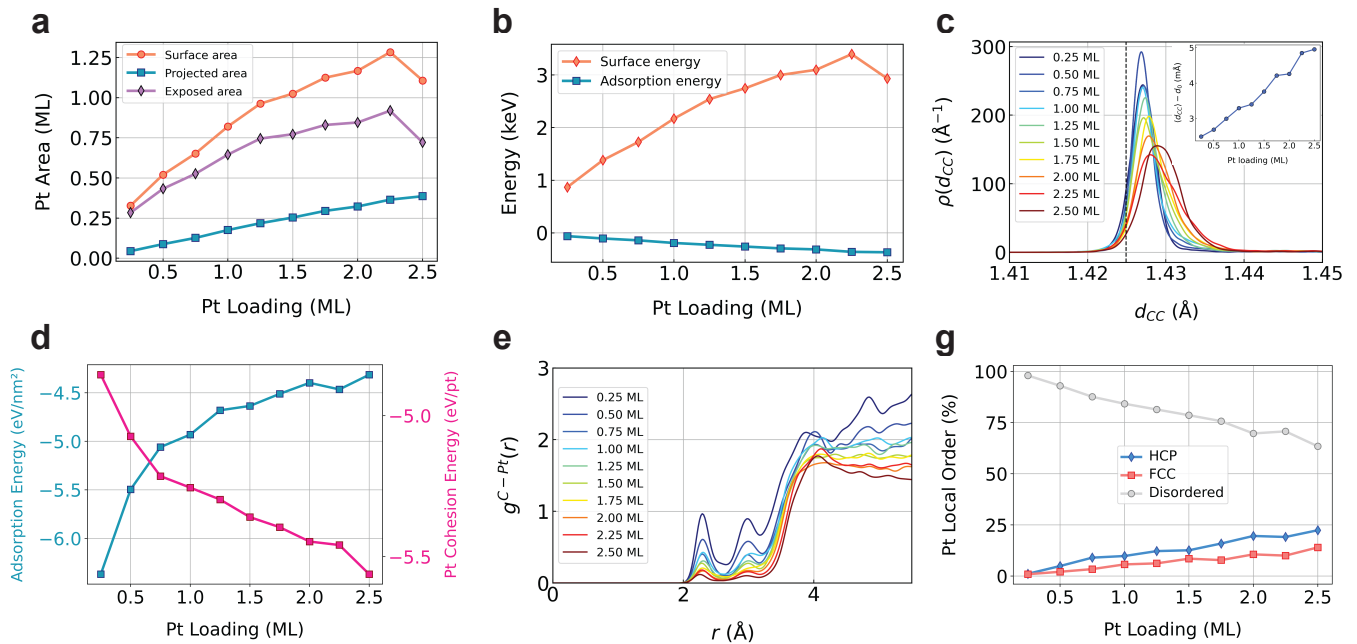}
    \caption{\textbf{Characterization of Pt Nanostructures on Graphene.} \textbf{a} Surface area ($A_{\text{surf}}$), projected area ($A_{\text{proj}}$), and exposed area ($A_{\text{expos}}$) of Pt clusters versus Pt loading. \textbf{b} Surface energy of isolated Pt structures and Pt adsorption energy on graphene, as a function of Pt loading. \textbf{c} Gaussian kernel density estimates of graphene C–C bond lengths, computed from the optimized structures using a 1.6 \AA\ neighbor cutoff and Scott’s bandwidth; the dashed line marks the DFT pristine bond length ($d_0 = 1.425$ \AA). Inset: Mean C–C bond‐length shift from pristine value (\(\bigl\langle d_{CC}\bigr\rangle - d_0\)) (in m\AA) versus Pt loading. \textbf{d} Pt-Pt cohesive energy (per Pt atom) and Pt-graphene adsorption energy (normalized by $A_{\text{proj}}$). \textbf{e} C–Pt radial distribution function for the optimized structures at various loadings. \textbf{f} Distribution of local atomic environments in Pt clusters (isolated from graphene), classified as FCC, HCP, or Disordered. }
    \label{fig:fig4}
\end{figure*}

\textbf{Energetic $\&$ Structural Characterization of Pt Clusters on Graphene.} Figure~\ref{fig:fig4} provides additional insights into the structural and energetic properties of Pt clusters on graphene, extending the morphological analysis from the previous section. We compute three characteristic areas: the surface area, $A_{\text{surf}}$ (for isolated Pt clusters, using Ovito’s Gaussian-density surface meshing modifier~\cite{krone2012fast, stukowski2009visualization}); the projected area, $A_{\mathrm{proj}}$ (discussed in the previous section); and the exposed area, defined as $A_{\mathrm{expos}} = A_{\mathrm{surf}} - A_{\mathrm{proj}}$. Figure~\ref{fig:fig4}a shows their evolution with Pt loading. $A_{\mathrm{proj}}$ increases nearly linearly, indicating consistent lateral domain growth, while $A_{\text{surf}}$ and $A_{\text{expos}}$ grow at a decelerating rate, peaking near $2.25$~ML and showing a slight rebound at $2.50$~ML. This rebound is attributed to the collective cluster coalescence observed at $2.50$~ML (Figures~\ref{fig:fig2}a, \ref{fig:fig3}, and S2f), which reduces the total surface area.

Figure~\ref{fig:fig4}b shows the evolution of surface ($E^{\text{surf}}_{\text{Pt}}$) and adsorption ($E^{\text{ads}}_{\text{Pt}}$) energies as a function of Pt loading, defined as follows:
\begin{align}
E^{\text{surf}}_{\text{Pt}} &= E^{\text{total}}_{\text{Pt}} - N_{\text{Pt}} \times E_{\text{Pt}}^{\text{bulk}},\\
E^{\text{ads}}_{\text{Pt-Gr}} &= E^{\text{total}}_{\text{Gr+Pt}} - E^{\text{total}}_{\text{Gr}} - E^{\text{total}}_{\text{Pt}}
\end{align}
where \(E^{\text{total}}_{\text{Pt}}\) is the total energy of the isolated Pt structure, \(N_{\text{Pt}}\) is the number of Pt atoms, \(E_{\text{Pt}}^{\text{bulk}}\) is the energy per atom in bulk FCC Pt, \(E^{\text{total}}_{\text{Gr+Pt}}\) is the total energy of the Pt/graphene system, and \(E^{\text{total}}_{\text{Gr}}\) is the total energy of relaxed bare graphene. 

The trend of $E_{\mathrm{Pt}}^{\mathrm{surf}}$ closely mirrors that of $A_{\mathrm{surf}}$, confirming that total surface energy scales primarily with surface area. Figure~\ref{fig:fig4}b further demonstrates that $E^{\text{ads}}_{\text{Pt}}$ becomes more negative and changes in a nearly linear fashion with Pt loading, paralleling $A_{\mathrm{proj}}$ and indicating larger Pt–graphene interactions at higher loadings. This is supported by Figure~\ref{fig:fig4}c, which shows Gaussian kernel density estimates of the graphene C–C bond length distributions from the optimized configurations; the inset plots the mean bond length deviation from pristine graphene. The progressive tensile strain introduced by increasing Pt loading induces rippling of the graphene sheet, as visually evident in Figure~\ref{fig:fig3}, and corroborated by the Raman spectra in Figure~S7 (detailed in the Experimental Section).

However, Figure~\ref{fig:fig4}d shows that the adsorption energy per unit area becomes less negative with increasing Pt loading, indicating weaker local Pt--graphene interfacial binding at higher coverages. These observations regarding Pt–graphene interactions imply that, while the local Pt--graphene bonds become weaker as the Pt loading increases, more Pt--graphene bonds are formed at a higher rate, leading to an overall stronger total adsorption energy between Pt and graphene and a correspondingly larger strain in graphene. According to the MLIP predictions in Figure~\ref{fig:fig4}d, the adsorption energies range from approximately $-6.4$~eV/nm$^2$ to $-4.3$~eV/nm$^2$, comparable to the binding energies reported for graphene/Pt(111)~\cite{meng2020step} ($\approx -4.1$~eV/nm$^2$), with the stronger adsorption energies at lower loadings attributed to the more under-coordinated nature of the smaller Pt nanoclusters. Overall, this range indicates relatively weak non-covalent binding between Pt nanostructures and graphene, yet remains stronger than the van der Waals (vdW) binding observed, for instance, between graphite layers ($\approx-2.0$~eV/nm\textsuperscript{2})~\cite{wei2015liquid}. 

Figure~\ref{fig:fig4}d further shows that as Pt--graphene binding weakens, the cohesive energy per Pt atom becomes more negative—reflecting enhanced Pt--Pt interactions. For reference, the bulk FCC Pt cohesive energy is calculated as $-5.91$ eV/atom by DFT and $-5.97$ eV/atom by our MLIP (experimental value $\approx-5.84$ eV/atom) \cite{kittel1996introduction}. This inverse relationship implies that, with increasing Pt loading, local Pt–graphene interactions weaken in favor of Pt–Pt cohesion, which progressively approaches that of the bulk. Figure~\ref{fig:fig4}e further corroborates these trends by showing the C–Pt radial distribution function, \(g^{\mathrm{C\text{-}Pt}}(r)\), normalized per Pt atom. Distinct maxima appear at approximately 2.3~\AA, 3.0~\AA, and 4.1~\AA. Our dilute-limit DFT calculations yield a C–Pt bond length of \(\approx2.1\)~\AA\ for an isolated Pt adatom at the bridge site on graphene, which closely aligns with the first peak predicted by the MLIP. However, at these higher loadings, this first peak shifts to \(\approx2.3\)~\AA, reflecting a slight elongation of the average C–Pt bond as Pt–Pt interactions begin to compete with Pt–graphene binding. Furthermore, the amplitude of \(g^{\mathrm{C\text{-}Pt}}(r)\) decreases with increasing loading, indicating that, on average, each Pt atom forms fewer short C–Pt contacts; in other words, a larger fraction of the Pt population becomes screened from the graphene lattice and agglomerates with neighboring Pt atoms away from graphene. Beyond \(\sim4\)~\AA, a long-range shoulder emerges above the first two peaks across all loadings, indicating that carbon atoms increasingly encounter Pt atoms at separations characteristic of multilayer metallic aggregates rather than isolated adatoms. Altogether, the \(g^{\mathrm{C\text{-}Pt}}(r)\) curves demonstrate that higher Pt loadings drive atoms to sacrifice strong, short C–Pt bonds in favor of increased Pt–Pt coordination, ultimately promoting the formation of more bulk-like Pt domains on graphene.

Lastly, Figure~\ref{fig:fig4}g employs the polyhedral template matching algorithm in Ovito~\cite{larsen2016robust, stukowski2009visualization}, with a root-mean-square deviation (RMSD) threshold of $0.12$, to classify the atomic environments within the Pt clusters based on coordination. Atomic environments are categorized as hexagonal close-packed (HCP), face-centered cubic (FCC), or "Disordered"—the latter denoting configurations that do not conform to any crystallographic template in Ovito’s reference set (including FCC, HCP, body-centered cubic, icosahedral, simple cubic, cubic diamond, or hexagonal diamond). Figure~\ref{fig:fig4}g shows that with increasing Pt loading, the fraction of disordered environments diminishes in favor of FCC and HCP orderings. This trend highlights the crystallographic heterogeneity of small Pt clusters, where high surface-to-volume ratios promote less-ordered arrangements; as clusters grow, surface effects diminish and crystalline FCC/HCP configurations emerge. Our DFT calculations indicate that the energy difference between bulk FCC and HCP stacking in Pt is $\approx53$~meV/atom (with FCC more stable), decreasing to $\approx31$~meV/atom for 6-layer [111]-like slabs and further to $\approx6$~meV/atom for 3-layer slabs. Consequently, in Pt nanostructures, the FCC and HCP phases become nearly degenerate, allowing the coexistence of multiple stacking orders at typical simulation temperatures (300~K \(\sim\) 26~meV).
\\

\textbf{Two‑Dimensional Pt Domains on Graphene.} Previous sections showed that gradual Pt deposition mimicking PVD under near-equilibrium conditions yields 3D nanoclusters as the ground‑state morphology. Here, we explore the potential stabilization of two‑dimensional (2D) Pt domains that may emerge under rapid, non‑equilibrium growth conditions. Such 2D structures (1–3 layers) have been experimentally realized on graphene via galvanic replacement of a sacrificial Cu layer~\cite{robertson2019atomic}. As a case study, we deposited a 2.00~ML Pt loading in a single batch on graphene to promote the stabilization of 2D domains. The resulting MD‑annealed structure (Figure~S3) reveals a stable 2D Pt film—up to 3 layers thick—with distinct monolayer (ML), bilayer (BL), and trilayer (TL) domains, consistent with experimental observations~\cite{robertson2019atomic}. Table~S1 compares the structural and energetic properties of the 2D film with its 3D nanocluster counterpart: the 2D film spans roughly \(3\times\) the projected area and exhibits \(\sim2\times\) the total adsorption energy of the nanoclusters, yet its adsorption energy per unit area is \(\sim30\;\%\) lower, indicating weaker local Pt–graphene binding. It also offers \(\sim1.7\times\) more Pt surface area, leading to a higher surface energy, and its formation energy is \(\approx0.22\) eV per Pt atom higher, marking lower thermodynamic stability relative to the nanocluster configuration.

To further investigate the preferred atomic coordination in 2D Pt domains, we performed DFT calculations for various free-standing BL models. Table~S2 and Figure~S4 summarize these structures classified by their in-plane lattice symmetry—square (sq) or hexagonal (hex)—and by their second-layer stacking orientation (top, center, or bridge). Among the square configurations, the sq--sq-top model yields a cohesive energy of $-4.86$~eV/atom, while sq--sq-center is slightly more stable at $-4.95$~eV/atom. These square-lattice domains, which have relatively higher energies, are considered metastable and may become kinetically trapped during rapid condensation on cooler substrates in the electron-beam PVD process (further discussed in the Experimental Section). In contrast, hexagonal configurations exhibit stronger cohesive energies of approximately $-5.19$~eV/atom across hex--hex-top, hex--hex-bridge, and hex--hex-center arrangements, with the hex--hex-center stacking being marginally more stable. These findings are consistent with the MLIP-driven MD results in Figure~S3, which show that BL regions adopt either hex--hex-top (AA) or hex--hex-bridge (AB) arrangements, while hex--hex-center stacking is absent due to its tendency to nucleate a third layer, transitioning toward an FCC- or HCP-like TL configuration.
\\

\begin{figure*}
    \centering
    \includegraphics[width=1.0\linewidth]{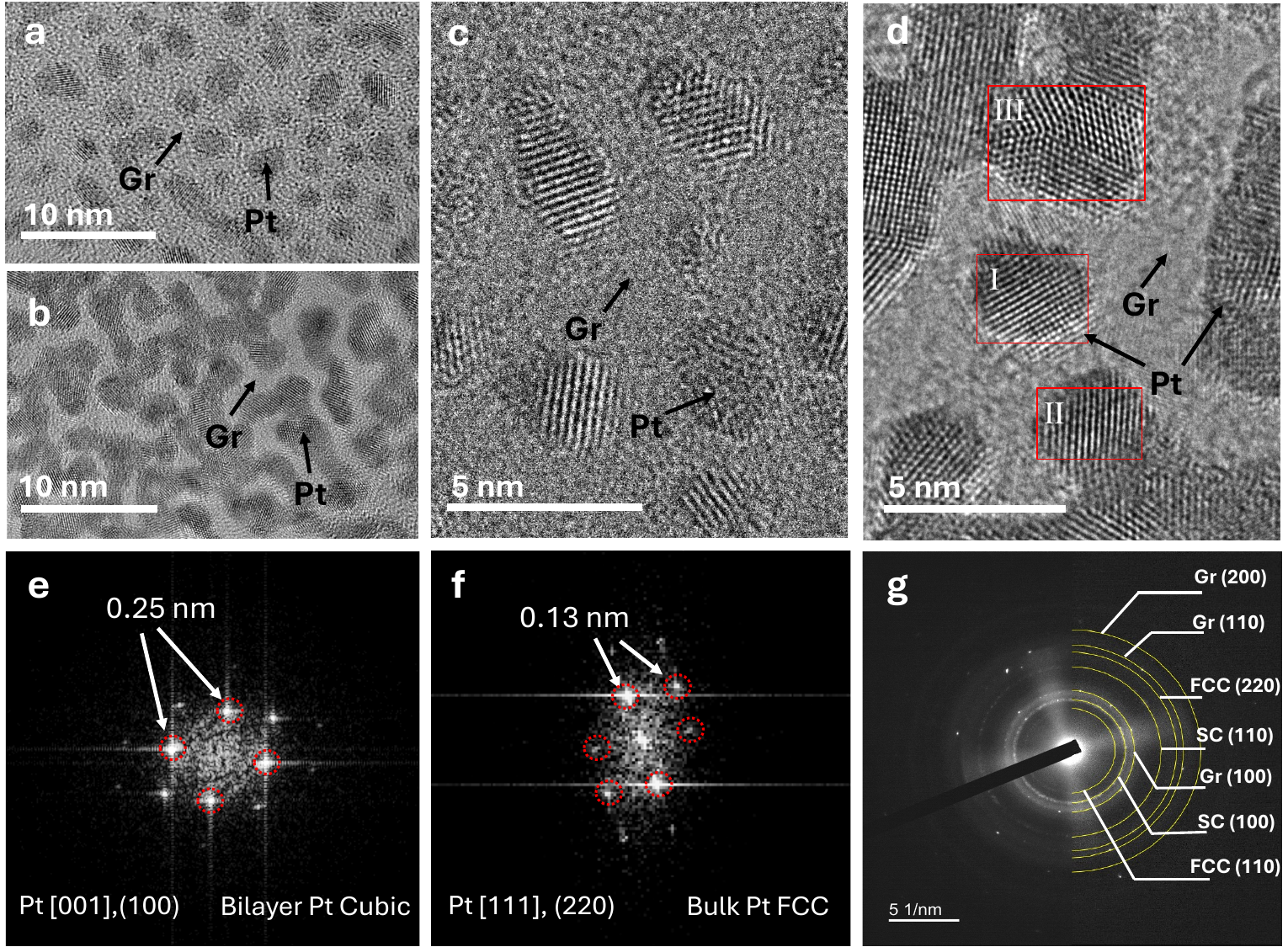}
    \caption{\textbf{TEM Characterization of Pt Nanostructures on Graphene.} TEM and HRTEM images at 1.5 minutes (\textbf{a}, \textbf{c}) and 3 minutes (\textbf{b}, \textbf{d}) of deposition are shown. \textbf{e} FFT analysis for region I, showing a cubic crystal structure viewed along the [001] axis. \textbf{f} FFT analysis for region III, showing an FCC crystal structure viewed along the [111] axis. \textbf{g} SAED pattern for Pt on graphene after 3 minutes of deposition.}
    \label{fig:fig5}
\end{figure*}

\textbf{Raman $\&$ TEM Characterization and Analysis.} Pt functionalization was performed using electron-beam PVD, as detailed in the "Methods" section. Prior to deposition, an outgassing step was conducted to remove native oxides and surface contaminants, ensuring a clean Pt deposition process. During deposition, the Pt film thickness was monitored using a crystal monitor to maintain precise control over film growth. Owing to the relatively fast deposition rate, two different deposition times—$1.5$ and $3.0$ minutes—were chosen, corresponding to approximately $0.4$ nm and $1.0$ nm of Pt on graphene, respectively. This yielded two distinct surface morphologies, enabling temporal tracking of film growth.

To investigate potential structural changes in graphene induced by Pt deposition, Raman characterization was conducted both before and after the deposition process. As shown in Figure~S7, both the $1.5$-minute and $3.0$-minute Pt depositions result in no significant shifts in the G (1580~cm\textsuperscript{-1}) or 2D (2700~cm\textsuperscript{-1}) peaks. Similarly, the intensity ratio (I$_{2D}$/I$_{G}$) remains approximately unchanged. This indicates that the graphene lattice remains essentially intact and that the Pt deposition process does not induce significant lattice disorder or substantial doping. However, it is worth noting that the G band exhibits slight broadening in both samples, with more pronounced broadening observed at the higher Pt loading. This behavior suggests localized charge interactions between Pt nanostructures and graphene~\cite{voggu2008effects}, corroborated by our MLIP-driven MD simulations, which show an increased Pt--graphene interface area and a stronger total adsorption energy with increasing Pt loading (see Figures~\ref{fig:fig4}a and~\ref{fig:fig4}b). Since the G band is highly sensitive to in-plane stretching of C--C bonds in sp$^2$ carbon, its broadening further suggests localized strain induced by Pt clustering on the graphene surface. This inference is supported by our MD simulations, which show that graphene develops ripples beneath the Pt clusters, with ripple amplitude increasing with Pt loading (Figure~\ref{fig:fig4}c). Meanwhile, the D peak at approximately 1350~cm\textsuperscript{-1} shows a small increase in intensity, indicating minor localized defects or lattice disruptions~\cite{canccado2024science}. Collectively, these findings highlight that Pt functionalization induces moderate local strain and charge interactions with graphene, while the overall integrity of the graphene film remains well preserved, with minimal disorder or defects. As a result, the advantageous electronic transport properties of graphene, which are essential for enhanced gas sensing, are retained.

High-resolution TEM (HRTEM) was employed to characterize the structural properties of the Pt/graphene structures. As shown in Figure~\ref{fig:fig5}a, after 1.5~minutes of deposition (\(\sim0.4\)~nm), the deposited Pt forms isolated nanoclusters dispersed on the graphene surface. Figure~\ref{fig:fig5}c presents a higher-magnification image of these Pt clusters, whose morphology aligns well with our MLIP predictions of small, nearly spherical Pt nanoclusters dispersed on graphene at lower Pt loadings (see Figure~\ref{fig:fig3}). After 3 min (\(\sim1.0\) nm) (Figure~\ref{fig:fig5}b), the Pt film develops larger, more closely spaced Pt domains via lateral coalescence of the initially formed smaller clusters—consistent with MLIP predictions (Figures~\ref{fig:fig2} and~\ref{fig:fig3}).

Three distinct Pt regions appear in the higher-magnification image of the \(1.0\)~nm loading (Figure~\ref{fig:fig5}d), highlighted by boxes~(I), (II), and (III). In region~(I), a relatively thin Pt domain lies on top of graphene. Fast Fourier transform (FFT) analysis of this region reveals a set of reflections consistent with a simple cubic (SC) crystal structure, viewed along the $[001]$ direction, exhibiting $(100)$ reflections at 0.25~nm (Figure~\ref{fig:fig5}e). According to previous reports~\cite{abdelhafiz2018epitaxial,robertson2019atomic}, such SC characteristics have been attributed to the formation of a Pt bilayer on graphene. However, our DFT calculations (Table~S2 and Figure~S4) indicate that the SC phase in Pt bilayer domains (sq-sq-top) is energetically less favorable than other in-plane hexagonal arrangements. The formation of these metastable SC thin domains may result from the rapid, non-equilibrium nature of electron-beam PVD, which condenses vapor-phase metal atoms onto a cooler substrate under conditions that may lack sufficient surface mobility to relax into more stable phases. In contrast, 2D Pt domains with higher mobility on graphene are expected to adopt in-plane hexagonal symmetry matching that of graphene—consistent with our MLIP predictions (see Figure~S3).

In region~(II), a thicker Pt domain is observed, exhibiting an FCC crystal structure viewed along the $[111]$ direction, with $(220)$ reflections at 0.13~nm, as confirmed by FFT analysis (Figure~\ref{fig:fig5}f). The formation of the FCC phase in this region is primarily attributed to the increased atomic mobility of Pt atoms as the film thickens~\cite{picone2014enhanced}. As Pt domains grow, the upper atoms become increasingly shielded from direct contact with graphene (no longer pinned by the Pt--graphene interface), reducing the substrate’s influence relative to the Pt--Pt cohesive interactions that begin to dominate within the Pt domain interior. Since FCC represents the global minimum-energy configuration for bulk platinum, these thicker regions tend to relax into the FCC phase. This is consistent with the MLIP-driven MD results shown in Figure~\ref{fig:fig4}g. Lastly, region~(III) exhibits FCC Pt domains with different orientations that merge into a larger cluster, in agreement with MLIP predictions regarding the cluster coalescence process (Figure~\ref{fig:fig2}).

To further investigate the crystal structure of the deposited Pt over a larger area, selected area electron diffraction (SAED) was performed following HRTEM and FFT analysis. As shown in Figure~\ref{fig:fig5}g, the diffraction rings confirm that Pt exhibits a polycrystalline structure composed of randomly oriented crystalline domains. Analysis of the interplanar spacings (Table~S3) reveals reflections corresponding to both FCC and SC phases of Pt. The SAED pattern (Figure~\ref{fig:fig5}g) shows reflections from FCC $(220)$ (at 0.13~nm), in agreement with the FFT analysis (Figure~\ref{fig:fig5}f). Additionally, reflections from SC $(100)$ (at 0.25~nm) and SC $(110)$ (at 0.15~nm) align with the FFT analysis (Figure~\ref{fig:fig5}e) and previous reports on bilayer Pt on graphene~\cite{robertson2019atomic}. For the DFT bilayer structures considered in Table~S2 and Figure~S4, the SC bilayer (sq-sq-top) exhibits a $(100)$ spacing of $\approx0.25$~nm, consistent with the experimental value. However, while the theoretical spacing for the $(110)$ plane is $0.25/\sqrt{2} \approx 0.178$~nm, we attribute the observed deviation to minor in-plane distortions. This is supported by the energetic favorability of hexagonally ordered Pt configurations, as noted in Table~S2. A small angular distortion of about 6 degrees could reduce the $(110)$ spacing from 0.178~nm to $\approx0.15$~nm. Furthermore, reflections at 0.31~nm are observed, close to the Pt FCC $(110)$ spacing of 0.28~nm. We suggest this slight discrepancy arises from a distortion of the Pt FCC lattice induced by the underlying hexagonal graphene substrate. Finally, the SAED pattern exhibits well-defined hexagonal graphene diffraction spots, further confirming that the structural integrity of graphene was maintained during the deposition process, without introducing significant defects, as also evidenced by Raman measurements. 
\\

\begin{figure*}
    \centering
    \includegraphics[
      width=1.0\textwidth,
      keepaspectratio
    ]{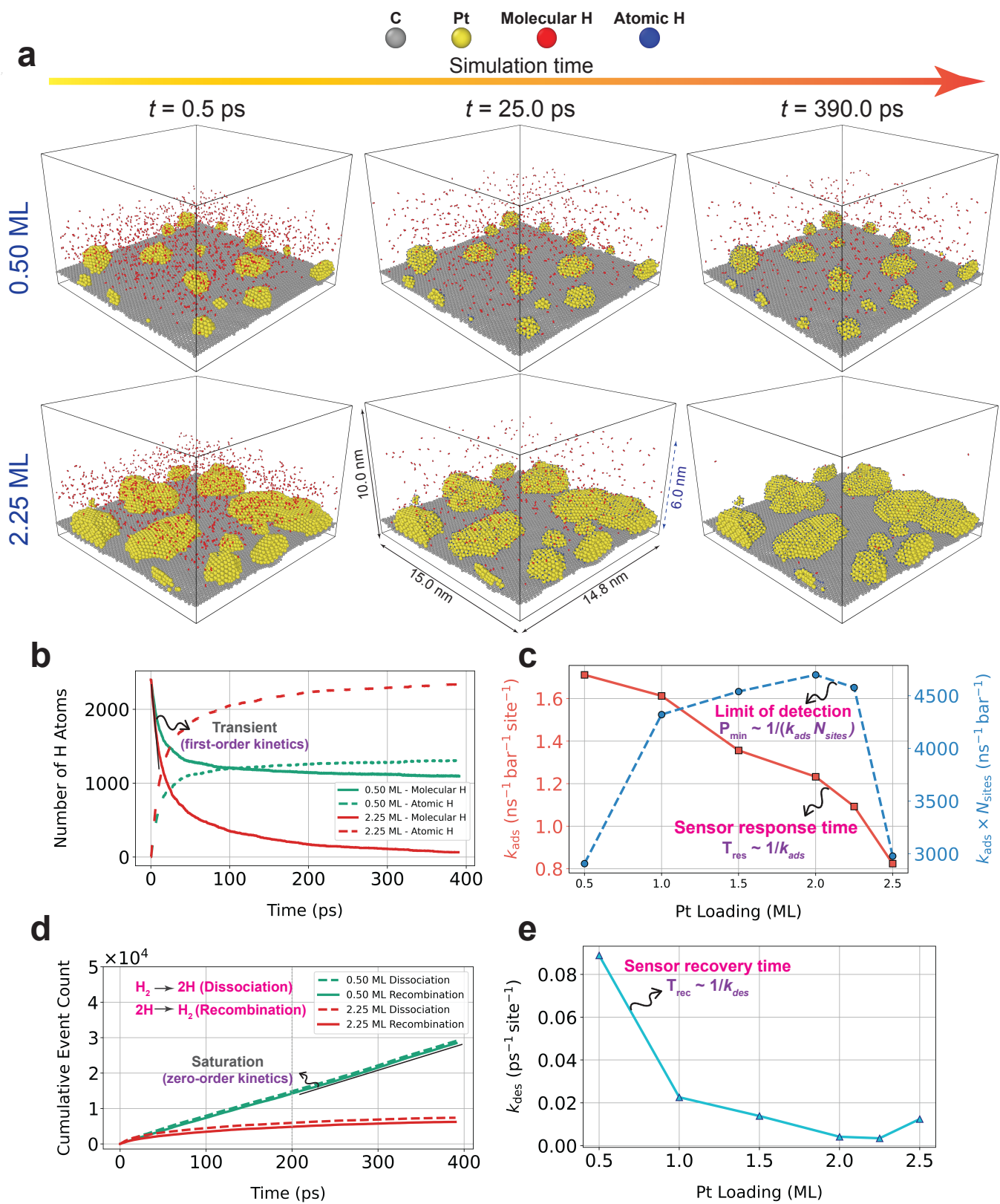}
    \caption{\textbf{H$_2$ Chemical Sensing on Pt-Functionalized Graphene.} \textbf{(a)} Time evolution of 1200 H$_2$ molecules interacting with optimized Pt/graphene structures at 2 loadings (0.50 ML and 2.25 ML), at 300 K with a 1 fs MD timestep. \textbf{(b)} Evolution of the number of H atoms in molecular/atomic state, determined by coordination (1 or 0) within a 1.0 \AA\ cutoff (for the 0.50 ML and 2.25 ML loadings). \textbf{(c)} Adsorption rate constant $k_{ads}$ (sticking coefficient) and the product \(k_{\mathrm{ads}}N_{\mathrm{sites}}\) plotted as functions of loading. $k_{ads}$ is calculated via Eq.~\eqref{eq:Ns_exp} from the effective rate constant $k_{\mathrm{eff}}$ obtained by linear fitting of Eq.~\eqref{eq:ln_fit} over the transient period (first 5–7~ps). \textbf{(d)} Cumulative counts of dissociation (H$_2$ $\rightarrow$ 2H) and recombination (2H $\rightarrow$ H$_2$) events over time (for the 0.50 ML and 2.25 ML loadings). \textbf{(e)} Desorption rate constant \(k_{\mathrm{des}}\), extracted from the equilibrium recombination flux \(R_{\mathrm{recom}}\) [Eq. \eqref{eq:kdes_final}], determined as the slope of a linear fit to the cumulative recombination events after 200 ps.}
    \label{fig:fig6}
\end{figure*}

\textbf{Hydrogen Reactive Sensing on Pt/Graphene Structures.} To investigate the hydrogen sensing behavior of Pt/graphene nanostructures, we conducted reactive MD simulations at $300$~K with $1200$ H$_2$ gas molecules, using the optimized Pt/graphene crystal structures at loadings of $0.50$~ML, $1.00$~ML, $1.50$~ML, $2.00$~ML, $2.25$~ML, and $2.50$~ML. The use of such high initial H$_2$ pressure increases the ensemble size, enabling accelerated reaction kinetics that remain computationally tractable within MD simulation timescales, as demonstrated in previous studies~\cite{vandermause2022active, owen2024atomistic}. In the initial configuration, we strategically position the H$_2$ molecules within a region from $5$ to $20$~\AA\ above the highest atom in the Pt/graphene structure, ensuring no direct initial contact. The simulation employs periodic boundaries in the lateral directions and incorporates a rigid wall approximately $6$ nm above the graphene sheet, serving as a reflecting barrier for the H$_2$ molecules. To suppress spurious velocity build‑up that may arise under the high H$_2$ background pressure, we periodically nullify the net linear momentum of the graphene substrate and the center-of-mass momentum of each Pt cluster. 

Figure~\ref{fig:fig6}a illustrates the dynamics of hydrogen interaction with Pt/graphene structures at two representative loadings (0.50~ML and 2.25~ML), with similar behavior observed across other loadings. Within 0.5~ps, the H$_2$ gas disperses uniformly throughout the simulation box. Over time, molecular hydrogen (H$_2$, shown in red) interacts primarily with the Pt clusters, undergoing dissociative chemisorption and converting into atomic hydrogen (H, shown in blue) on the Pt surfaces. Notably, the simulations indicate that hydrogen exhibits negligible direct interaction with pristine graphene. Figure~S5 shows the H--Pt and H--C radial distribution functions (\(g^{\mathrm{H\text{-}Pt}}(r)\) and \(g^{\mathrm{H\text{-}C}}(r)\)), obtained from the relaxed final MD snapshots at different Pt loadings. \(g^{\mathrm{H\text{-}Pt}}(r)\) displays a pronounced first peak at \(\approx 1.75\)~\AA, consistent with previous reports~\cite{hanh2014first}, whereas \(g^{\mathrm{H\text{-}C}}(r)\) lacks a defined peak at the typical H--C covalent bond length (\(\approx 1.1\)~\AA) and remains negligible for \(r < 2.25\)~\AA, indicating the absence of H--C chemical bonding and the presence of only van der Waals interactions with graphene. 

Note that our Raman analysis (Figure~S7) reveals minimal defect levels in graphene, justifying the assumption of pristine graphene in our simulations. Under standard conditions, unfunctionalized pristine graphene is known to be chemically inert to H$_2$, with a high dissociation barrier (\(\ge3.3\)~eV)~\cite{miura2003first}. However, the absence of H--graphene chemical interaction even after Pt functionalization—which facilitates H$_2$ dissociation and dopes graphene—suggests that H spillover from Pt clusters to pristine graphene is negligible, a behavior also observed visually in our MD simulations (Figure~\ref{fig:fig6}a). The DFT binding energy of a single H adatom on graphene (at the top site) is $-0.9$~eV, in agreement with previous studies, and is slightly overestimated by our MLIP at $-1.19$~eV~\cite{dumi2022binding}. These results indicate that atomic H adsorption on graphene is exothermic according to both MLIP and DFT, ruling out any potential MLIP bias that might underestimate the energetic favorability of H adsorption on graphene and thereby artificially suppress H spillover. In fact, the absence of H spillover can be attributed to the relatively weak Pt–graphene binding, as confirmed by our MLIP predictions and Raman spectroscopy (Figures~\ref{fig:fig4}d and S7). This weak Pt--graphene interaction increases the electron density at Pt surface sites, thereby strengthening H--Pt bonds (Figure~\ref{fig:fig6}d), elevating the H desorption barrier, and impeding hydrogen migration onto the graphene lattice. Prior studies support this hypothesis, demonstrating that weaker metal–substrate interactions hinder H spillover~\cite{sihag2019dft, yeh2025dft, sihag2023defects}. Other studies also show that the H spillover barrier from Pt clusters to pristine graphite is very high ($> 50$ kcal/mol)~\cite{psofogiannakis2009dft}. Although hydrogen primarily adsorbs on Pt, these adsorption events can still induce bonding perturbations detectable at the Pt--graphene interface, providing a sensing channel distinct from a direct H--graphene interaction, as detailed in the next section.

Figure~\ref{fig:fig6}b shows the time evolution of atomic and molecular hydrogen for Pt loadings of 0.50~ML and 2.25~ML as representative cases, with all other loadings exhibiting similar behavior. Each hydrogen atom is classified as either atomic (H) or molecular (H$_2$) based on its coordination number (considering only neighboring H atoms) within a 1.0\,\AA\ cutoff. Figure~\ref{fig:fig6}b displays an initial, approximately exponential increase in atomic hydrogen accompanied by a corresponding decay in molecular hydrogen during the transient response over the first 0–10~ps, indicating first-order kinetics in this interval. Under Langmuir kinetics with second‐order dissociative adsorption and recombinative desorption of hydrogen, the fractional surface coverage $\theta = N^{(s)}_{\mathrm{H}}/N_{\mathrm{sites}}$ evolves according to
\begin{equation}
\frac{d\theta}{dt}
= k_{\mathrm{ads}}\,P_{\mathrm{H}_2}\,(1-\theta)^2
- k_{\mathrm{des}}\,\theta^2
\label{eq:langmuir}
\end{equation}
where $N^{(s)}_{\mathrm{H}}$ is the number of chemisorbed H atoms, $k_{\mathrm{ads}}$ and $k_{\mathrm{des}}$ are the dissociative adsorption and recombinative desorption rate constants respectively, and $P_{\mathrm{H_2}} = N^{(g)}_{\mathrm{H_2}}\,k_{\mathrm{B}}\,T/V$ denotes the instantaneous gas pressure in the NVT simulation cell. The number of surface sites is approximated by \(N_{\mathrm{sites}} = A_{\mathrm{surf}}/(\tfrac{\sqrt{3}}{2}\,d_{\mathrm{nn}}^2)\), where \(d_{\mathrm{nn}} = 2.81\)\,\AA, yielding a surface site density of \(\approx1.46 \times 10^{15}\,\mathrm{sites/cm}^2\)~\cite{sharma2018microkinetic}.

During the transient period, the sensor operates in the low‐coverage regime (\(\theta \ll 1\)), and our MD trajectories exhibit an approximately exponential increase in \(N_H^{(s)}\), both implying first‐order kinetics. We therefore linearize the full Langmuir model [Eq.~\eqref{eq:langmuir}] by neglecting higher‐order terms \(O(\theta^2)\). Under these conditions, the surface population evolves as
\begin{equation}
\frac{d\theta}{dt}
\approx k_{\mathrm{ads}}P_0
\Bigl[1 - (b+2)\,\theta\Bigr]
\label{eq:low_coverage_expansion}
\end{equation}
where \(b = N_{\mathrm{sites}}/(2N^{(g)}_{\mathrm{H}_{2}}(0))\) and \(P_{0} = N^{(g)}_{\mathrm{H}_{2}}(0)\,k_{\mathrm{B}}T/V\) (detailed derivation is provided in the SI). Physically, \(b\) measures how many adsorption sites there are per initial gas molecule, and the product \(b\,P_0\) is the rate at which the driving gas pressure falls from $P_0$ as coverage \(\theta\) increases ($P(t) \;=\; P_0(1 - b\,\theta(t))$). The sum \(b+2\) in Eq.~\eqref{eq:low_coverage_expansion} thus embodies two independent first‐order retardation effects on the adsorption flux from its initial $k_{\mathrm{ads}}P_0$: the “2” reflects that each dissociative event consumes two empty sites—reducing the flux by \(2\,k_{\mathrm{ads}}P_0\,\theta\)—while the term \(b\) captures gas‐phase depletion—reducing the flux by \(b\,k_{\mathrm{ads}}P_0\,\theta\). 

Solving this first‐order ordinary differential equation [Eq.~\eqref{eq:low_coverage_expansion}] yields
\begin{equation}
N_H^{(s)}(t)=\frac{N_{\mathrm{sites}}}{b+2}\bigl[1-e^{-k_{\mathrm{eff}}t}\bigr],\quad k_{\mathrm{eff}}=(b+2)\,k_{\mathrm{ads}}P_0
\label{eq:Ns_exp}
\end{equation}
(valid for the transient period, with low‐coverage \(\theta\ll1\)). \(k_{\mathrm{eff}}\) is obtained by fitting the log‐residual 
\begin{equation}
\ln\!\Bigl[1 - \tfrac{b+2}{N_{\mathrm{sites}}}\,N_H^{(s)}(t)\Bigr]
= -\,k_{\mathrm{eff}}\,t
\label{eq:ln_fit}
\end{equation}
whose slope yields \(-k_{\mathrm{eff}}\). Physically, \(k_{\mathrm{eff}}\) represents the overall first‐order rate constant of the sensor during its initial gas exposure under NVT simulation, accounting for both the slowdown due to dissociative site blocking (\(2\,k_{\mathrm{ads}}P_0\)) and gas‐phase depletion (\(b\,k_{\mathrm{ads}}P_0\)).

Figure~\ref{fig:fig6}c presents \(k_{\mathrm{ads}}\) as a function of Pt loading, extracted from the initial 5--7~ps of simulation data sampled at 5~fs intervals. Within this brief interval, $\theta$ increases by only 15--21\;\% across different loadings; thus, $\theta^2\lesssim 0.04$, validating the assumption made during the transient period to neglect the $\mathcal{O}(\theta^2)$ terms. From Figure~\ref{fig:fig6}c, it is evident that $k_{\mathrm{ads}}$ decreases monotonically with increasing Pt loading, reflecting the progressive loss of under-coordinated edge and corner sites in favor of terrace sites (Figure~S2h). The zero-coverage sticking probability \(S_0\) is a standard descriptor of surface reactivity, representing the fraction of incident molecules that undergo adsorption in the low-coverage limit. For dissociative adsorption of H\(_2\), the coverage-dependent sticking probability follows \(S(\theta) = S_0(1 - \theta)^2\). \(S_0\) is directly related to the intrinsic adsorption rate constant \(k_{\mathrm{ads}}\) by
\[
S_0 = \frac{\sqrt{2\,m_{\mathrm{H}_2}\,k_B\,T}}{A_{\rm site}}\;\frac{k_{\mathrm{ads}}}{2}
\]
where \(A_{\rm site}\) is the footprint of a single Pt(111) adsorption site, and the factor of \(1/2\) converts \(k_{\mathrm{ads}}\) from a per-H-atom (site) rate to a per-H\(_2\)-molecule sticking rate. $S_0$ follows the same trend as $k_{\mathrm{ads}}$, with values ranging from 0.66 at 0.50~ML down to 0.32 at 2.50~ML loading, in agreement with literature~\cite{kraus2017dynamics, ghassemi2019assessment}.

\medskip
The \emph{response time}, $T_{\mathrm{res}}$, is conventionally defined as the time required to achieve 90\;\% of the equilibrium sensor signal (i.e., 90\;\% equilibrium surface coverage). Rather than analyzing $T_{\mathrm{res}}$ directly, we quantify the responsiveness kinetics via the first-order time constant obtained from the low-coverage transient period. Under realistic operating conditions, the device is exposed to a continuously replenished gas stream ($P=\mathrm{constant}$); thus, gas-phase depletion is negligible, and the coefficient $b$ vanishes ($b = N_{\mathrm{sites}}/(2N^{(g)}_{\mathrm{H}_2}) \sim N_{\mathrm{sites}}/\infty \sim 0$). With $b=0$, the time constant reduces to:
\begin{equation}
\tau_{\mathrm{res}}
=\frac{1}{k_{\mathrm{eff}}}
=\frac{1}{2k_{\mathrm{ads}}P_{0}}
\label{eq:tau_res_constP}
\end{equation}
showing that $\tau_{\mathrm{res}}$ is inversely proportional to $k_{\mathrm{ads}}$. Because $k_{\mathrm{ads}}$ decreases monotonically with increasing Pt loading (Figure~\ref{fig:fig6}c), $\tau_{\mathrm{res}}$ exhibits a corresponding increase throughout the series. Furthermore, since $\tau_{\mathrm{res}}\propto P_{0}^{-1}$, reducing the hydrogen partial pressure from the simulated value $P_{0}\approx37.3$ bar to an experimentally relevant level $P_{\mathrm{exp}}\approx10^{-6}$ bar (1~ppm) lengthens the response time by a factor
$P_{0}/P_{\mathrm{exp}}\approx3.7\times10^{7}$—a shift of 7–8 orders of magnitude into an experimentally relevant range where $\tau_{\mathrm{res}}$ ranges from about $0.3$ ms ($0.50$ ML) to $0.6$ ms ($2.50$ ML).

\medskip
\emph{The limit of detection}, LOD, is another essential figure of merit for gas sensors, representing the lowest gas partial pressure distinguishable above the noise floor ($\sigma_{\mathrm{noise}}$). Assuming each chemisorbed H atom contributes a conductance increment \(\Delta G_{1}\), the short-time expansion of Eq.~\eqref{eq:Ns_exp}, valid for interrogation times \(t_{\mathrm{int}} \ll 1/k_{\mathrm{eff}}\), yields
\begin{equation}
N_H^{(s)}(t_{\mathrm{int}})
\approx \frac{N_{\mathrm{sites}}}{b+2}\,k_{\mathrm{eff}}\,t_{\mathrm{int}}
= N_{\mathrm{sites}}\,k_{\mathrm{ads}}\,P_0\,t_{\mathrm{int}}
\end{equation}
Consequently, the corresponding conductance change is
\begin{equation}
\Delta G \approx \Delta G_{1}\,N_H^{(s)}\bigl(t_{\mathrm{int}}\bigr)
= \Delta G_{1}\,N_{\mathrm{sites}}\,k_{\mathrm{ads}}\,P_0\,t_{\mathrm{int}}\
\end{equation}
Requiring \(\Delta G \ge \sigma_{\mathrm{noise}}\) gives the minimum detectable pressure
\begin{equation}\label{eq:Pmin_final}
P_{\min}
\approx \frac{\sigma_{\mathrm{noise}}}
     {\Delta G_{1}\,N_{\mathrm{sites}}\,k_{\mathrm{ads}}\,t_{\mathrm{int}}}\,
\end{equation}

Figure~\ref{fig:fig6}c illustrates the product \(k_{\mathrm{ads}}N_{\mathrm{sites}}\)—the metal loading–dependent factor of \(P_{\min}\)—highlighting the competition between the increasing \(N_{\mathrm{sites}}\) and decreasing \(k_{\mathrm{ads}}\) with Pt loading. The observed trend demonstrates that the growth in Pt surface area (and, in turn, \(N_{\mathrm{sites}}\)) outweighs the intrinsic reduction in sticking kinetics as Pt loading increases and surface morphology evolves toward more coordinated sites. The product \(k_{\mathrm{ads}}N_{\mathrm{sites}}\) peaks at 2.00\,ML \(\bigl(4.7\times10^{3}\,\mathrm{ns^{-1}\,bar^{-1}}\bigr)\), delivering an LOD that is \(\sim\!1.6\times\) lower than at the 0.50~ML and 2.50~ML loadings. Figures~\ref{fig:fig4}a–b confirm that increasing Pt loading up to $\sim2.25$~ML enhances both surface area and surface energy; beyond this threshold, stronger edge coalescence reduces these metrics. This phenomenon, combined with the monotonically decreasing \(k_{\mathrm{ads}}\) as Pt loading increases, explains the substantial decline in \(k_{\mathrm{ads}}N_{\mathrm{sites}}\) at 2.50~ML, where both \(A_{\mathrm{surf}}\) and \(k_{\mathrm{ads}}\) decrease relative to the 2.25~ML loading, despite the higher Pt mass. Overall, these observations reveal the existence of an optimum intermediate Pt loading that minimizes the LOD.
 
\medskip
\emph{The recovery time}, \(T_{\mathrm{rec}}\), the third figure of merit we consider, is defined as the time required to desorb 90\;\% of the adsorbed gas species. \(T_{\mathrm{rec}}\) is inversely proportional to the per-site desorption rate constant \(k_{\mathrm{des}}\) (s\(^{-1}\) per site). Analysis of the post‑equilibration segment of our MD trajectories (after \(\sim\!200\)~ps; Figure~\ref{fig:fig6}d) reveals nearly constant dissociation and recombination rates, indicative of zeroth‑order desorption (\(n = 0\)) in the saturation regime. In this regime, \(\mathrm{H}_2\) adsorption–desorption kinetics are independent of both gas pressure and surface coverage \(\theta\); consequently, the per‑site desorption rate equals the per‑site desorption rate constant. According to the Polanyi–Wigner relation for \(n=0\) \cite{schroeder2002temperature}, the per-site desorption rate is given by
\[
r_{\mathrm{des}} = k_{\mathrm{des}}\,\theta^0 = k_{\mathrm{des}}
\]
\(k_{\mathrm{des}}\) would nominally follow an Arrhenius form, \(k_{\mathrm{des}} = \nu\exp\bigl(-|E_{\mathrm{ads}}|/(k_{\mathrm{B}}T)\bigr)\), but instead we extract \(k_{\mathrm{des}}\) directly from the MD‐measured recombination flux \(R_{\mathrm{recom}}\) (H\(_2\to2\mathrm{H}\) events/ps). Each recombination event consumes two surface H atoms, giving a total H‐atom removal rate of \(2R_{\mathrm{recom}}\). Dividing by the number of adsorption sites, \(N_{\mathrm{sites}}\), yields
\begin{equation}\label{eq:kdes_final}
k_{\mathrm{des}} = \frac{2\,R_{\mathrm{recom}}}{N_{\mathrm{sites}}}
\end{equation}
We focus on H\(_2\) recombination as the dominant desorption channel because direct atomic H desorption would require overcoming a high adsorption energy ($\approx2.87$~eV per H atom, as predicted by our MLIP and consistent with literature~\cite{vurdu2018adsorption,shi2017adsorption}), rendering that pathway negligible. To determine \(R_{\mathrm{recom}}\), we track the cumulative number of recombination events (\(2\mathrm{H}\to\mathrm{H}_2\)) in 50 fs time bins after a 200 ps equilibration period and compute the slope of the resulting linear cumulative‐event curve. The intrinsic constant \(k_{\mathrm{des}}\) (s\(^{-1}\)) quantifies the desorption frequency of a single site and thus provides a consistent basis for comparing \(T_{\mathrm{rec}}\) across different Pt loadings. Under practical conditions, the surface may be only partially covered, and thus the desorption rate depends on the coverage; however, the elementary rate constant \(k_{\mathrm{des}}\) remains unchanged.

Figure~\ref{fig:fig6}e shows that the MLIP-predicted desorption rate constant \(k_{\mathrm{des}}\) ranges from 0.003 to 0.089~ps\(^{-1}\). Assuming a typical attempt frequency of \(\nu = 10^{13}\,\mathrm{s}^{-1}\), these values correspond to an effective desorption barrier of $\sim0.12–0.21$~eV, comparable to experimental values on Pt(111) surfaces of 0.21–0.24~eV~\cite{vandermause2022active,christmann1976adsorption}. Moreover, \(k_{\mathrm{des}}\) decreases with increasing Pt loading, reaching a minimum at 2.00~ML and 2.25~ML before rising slightly at 2.50~ML. This behavior reflects the morphological evolution of the Pt structures, where lower loadings exhibit a greater fraction of under‐coordinated edge and corner sites (Figure~S2h), thereby enhancing desorption kinetics~\cite{starr2008large}. Our MD simulations thus show that smaller Pt clusters—rich in edge and corner sites—exhibit higher \(k_{\mathrm{des}}\) and correspondingly shorter recovery times. Consequently, larger clusters require higher peak desorption temperatures; under a fixed heating rate, this translates directly into longer times to remove the adsorbates and recover the bare surface.
\\

\begin{figure}
    \centering
    \includegraphics[width=0.49\textwidth]{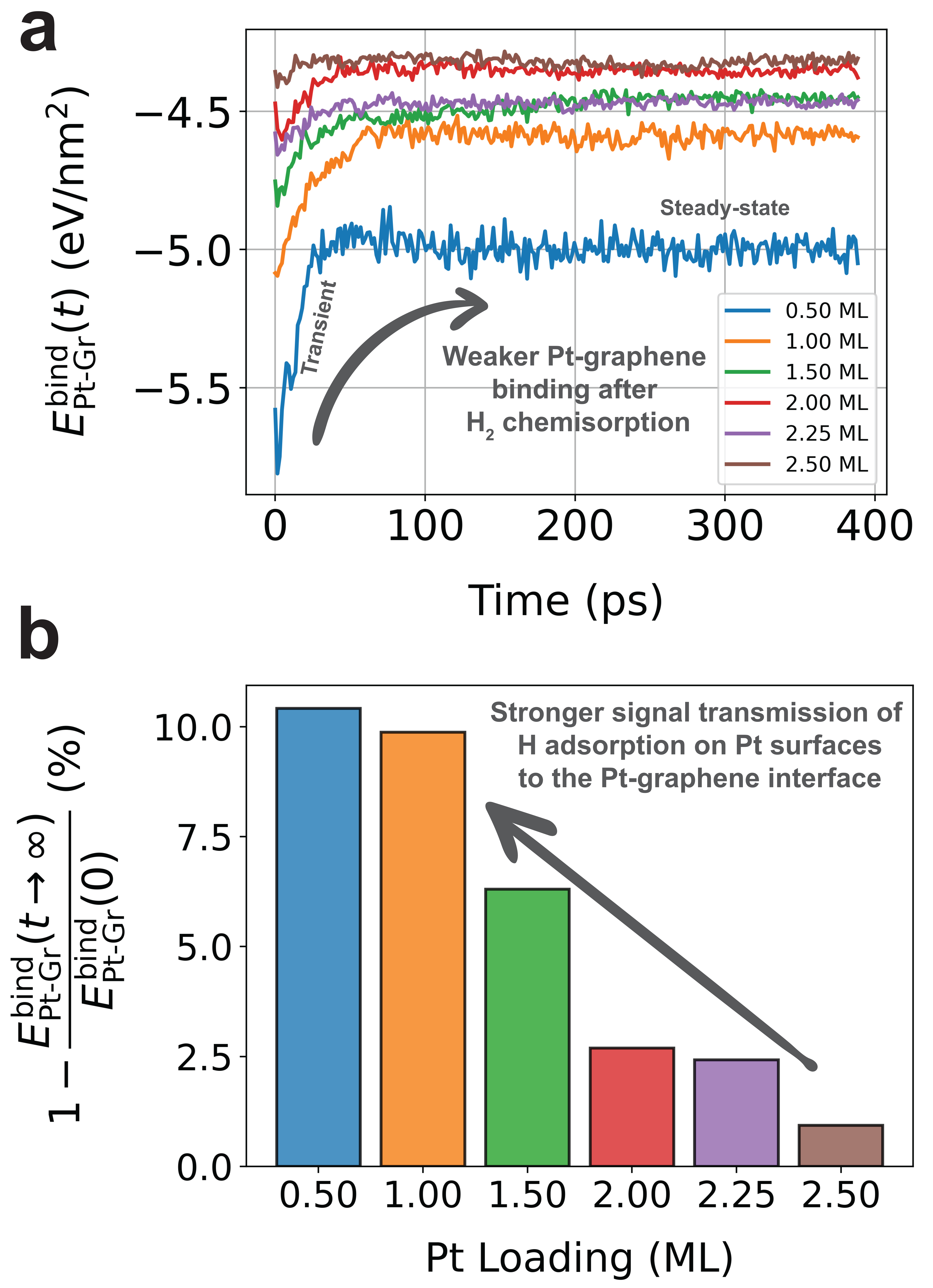}
    \caption{\textbf{Time Evolution of the Pt–Graphene Binding Energy.} 
    \textbf{(a)} Time-dependent binding energy for different loadings, sampled at 1.5~ps intervals during MD reactive sensing simulations with H$_2$ gas. \textbf{(b)} Binding-energy–based transduction sensitivity used to quantify the relative attenuation of Pt–graphene binding after H adsorption on Pt surfaces, as a function of Pt loading.}
    \label{fig:binding_energy}
\end{figure}

\textbf{Signal Transduction at the Pt–Graphene Interface.} Since dissociated hydrogen chemisorbs exclusively on Pt nanostructures—without chemically interacting with pristine graphene, as our reactive MD simulations show (Figure~S5b)—it is instructive to examine how hydrogen adsorption perturbs the Pt–graphene interface. To quantify these perturbations, we extract snapshots from the reactive MD trajectories with H\textsubscript{2} gas, remove all hydrogen atoms \emph{in situ}, and compute the instantaneous Pt–graphene binding energy (without further relaxation of the remaining atoms)
\begin{equation}
E^{\mathrm{bind}}_{\mathrm{Pt\text{-}Gr}}(t)
=
E^{\mathrm{tot}}_{\mathrm{Gr+Pt}}(t)
-
E^{\mathrm{tot}}_{\mathrm{Gr}}(t)
-
E^{\mathrm{tot}}_{\mathrm{Pt}}(t)
\label{eq:Ebind_time}
\end{equation}
where \(E^{\mathrm{tot}}_{\mathrm{Gr+Pt}}(t)\) is the total energy of the hydrogen-free Pt–graphene configuration at time \(t\), \(E^{\mathrm{tot}}_{\mathrm{Gr}}(t)\) is the total energy of the graphene sheet alone, and \(E^{\mathrm{tot}}_{\mathrm{Pt}}(t)\) is the total energy of the Pt structure alone.

Equation~\eqref{eq:Ebind_time} allows us to directly compare the evolution of Pt–graphene binding. It captures all many-body interactions between Pt and graphene at the hydrogen-induced geometry while automatically excluding H–Pt, H–Gr, and higher-order H–Pt–Gr contributions, thereby obviating the complex decomposition required by subtraction-based schemes that compute the Pt–H/graphene binding energy and then remove H–Gr and higher-order H–Pt–Gr terms to compare with the bare Pt–graphene binding or across different chemisorbed H loadings.

Figure~\ref{fig:binding_energy}a shows that the magnitude of \(E^{\mathrm{bind}}_{\mathrm{Pt\text{-}Gr}}(t)\) decays approximately exponentially after H$_2$ exposure before reaching a plateau once the cluster is saturated, indicating a systematic weakening of Pt–graphene adhesion as the Pt nanostructure becomes progressively hydrogenated. Although pristine graphene is chemically inert toward H$_2$, the hybrid Pt/graphene sensor responds immediately through this attenuation of interfacial binding. 

To quantify the relative attenuation of Pt–graphene binding across different Pt loadings, we define a dimensionless transduction sensitivity
\begin{equation}
\label{eq:sensing_response}
S_{\mathrm{Pt\text{-}Gr}}
\;=\;
1
-
\frac{%
  E^{\mathrm{bind}}_{\mathrm{Pt\text{-}Gr}}(t\!\to\!\infty)%
}{%
  E^{\mathrm{bind}}_{\mathrm{Pt\text{-}Gr}}(0)%
}
\times 100\;\%
\end{equation}
where \(E^{\mathrm{bind}}_{\mathrm{Pt\text{-}Gr}}(t\!\to\!\infty)\) is estimated as the average binding energy over the final 40 considered MD frames, and \(E^{\mathrm{bind}}_{\mathrm{Pt\text{-}Gr}}(0)\) is the initial binding energy (before hydrogen interaction). A larger positive \(S_{\mathrm{Pt\text{-}Gr}}\) indicates a stronger relative attenuation of Pt–graphene binding upon hydrogenation.

Figure~\ref{fig:binding_energy}b illustrates that smaller Pt clusters exhibit the largest relative weakening of the Pt-graphene binding after hydrogenation, with \(S_{\mathrm{Pt\text{-}Gr}}\) decreasing monotonically as Pt loading grows. This arises from $(i)$ the intrinsically weaker bare adhesion of small clusters to graphene (Figure~\ref{fig:fig4}b)—so that even a modest absolute loss \(\Delta E^{\mathrm{bind}}_{\mathrm{Pt\text{-}Gr}}\) constitutes a large fraction of the initial binding energy, boosting \(S_{\mathrm{Pt\text{-}Gr}}\), $(ii)$ the higher surface-to-volume ratio of small clusters, which makes H-induced attenuation of the interfacial binding more pronounced, and $(iii)$ the lower electronic screening by inner Pt layers in smaller clusters compared to larger ones. Therefore, smaller Pt loadings (yielding smaller dispersed clusters) are expected to serve as more responsive motifs for H$_2$ signal transduction at the Pt–graphene interface.
\\

\begin{table*}
\centering
\caption{DFT‐calculated Bader charges (in units of $10^{-3}e$) and adsorption energies (eV) for various Pt nanostructures on graphene, before and after hydrogenation. \(\Delta Q_{\mathrm{Gr}}^{\mathrm{bare/H}}\) gives the electron density transferred to graphene in the bare/H cases, corresponding to the unhydrogenated (Pt only) and hydrogenated (Pt+H) configurations. \(\Delta Q_{\mathrm{H}}/N_{\mathrm{H}}\) is the electron density transferred to hydrogen atoms, normalized per H atom. \(E_{\mathrm{ads}}^{\mathrm{bare/H}}\) lists the adsorption energy on graphene in the bare/H cases.}
\label{tab:charge_analysis}
\small
\begin{tabularx}{\textwidth}{
>{\raggedright\arraybackslash}p{0.35\textwidth}  
>{\centering\arraybackslash}X                    
>{\centering\arraybackslash}X                    
>{\centering\arraybackslash}X}                   
\hline
\textbf{Pt structure (supercell)} &
\(\Delta Q_{\mathrm{Gr}}^{\text{bare/H}}\) &
\(\Delta Q_{\mathrm{H}}/N_{\mathrm{H}}\) &
\(E_{\mathrm{ads}}^{\text{bare/H}}\) (eV) \\
\hline
Pt\(_1\) adatom (5\(\times\)5)          & \(+5.4/-139.0\) & \(+135.3\)  & \(-1.74/-0.98\) \\[4pt]
Pt\(_9\) nanocluster (5\(\times\)5)     & \(+55.9/+21.6\) & \(+8.9\)  & \(-2.39/-1.66\) \\[4pt]
Pt\(_{17}\) nanocluster (7\(\times\)7)  & \(+85.9/+53.4\) & \(+7.4\) & \(-3.04/-2.84\) \\[4pt]
Pt(111) 4-layer slab (2\(\times\)2)  & \(-49.7/-46.7\)  & \(-28.5\) & \(-0.31/-0.31\) \\
\hline
\end{tabularx}
\end{table*}

\textbf{Modulation of Charge Transfer at Pt–Graphene Interfaces by Hydrogen Adsorption.} To analyze the charge transfer between Pt nanostructures and graphene, and to assess the influence of Pt morphology and hydrogen adsorption on this transfer, we performed DFT calculations on a series of Pt structures supported on graphene, both in their bare form and after hydrogen adsorption on the Pt surfaces. Table~\ref{tab:charge_analysis} presents the principal findings from Bader charge analysis and adsorption energy calculations for 4 representative Pt/graphene structures: a single Pt adatom (Pt$_1$), two nanoclusters (Pt$_9$ and Pt$_{17}$), and a 4-layer Pt(111) slab (as modeled in~\cite{khomyakov2009first}). The slab model serves as a prototypical representation of a continuous, bulk-like Pt film interfaced with graphene.

As shown in Table~\ref{tab:charge_analysis}, the bare Pt$_1$, Pt$_9$, and Pt$_{17}$ adsorbates all donate electrons to graphene ($n$-type doping), with integrated charge transfers of \(+5.4\), \(+55.9\), and \(+85.9 \times 10^{-3}\,e\), respectively. After normalizing by the number of Pt atoms, each Pt structure contributes nearly the same amount—\(\sim\!5{-}6 \times 10^{-3}\,e\) per Pt—suggesting that the per-atom donation is roughly size-independent. The 4-layer extended Pt(111) slab reverses the direction of charge transfer: the slab withdraws about \(49.7 \times 10^{-3}\,e\) from graphene, or about \(-4.1 \times 10^{-3}\,e\) per Pt atom. Although the magnitude of charge transfer per Pt atom remains comparable to the finite Pt structures (further supporting the approximately size-independent charge donation per Pt atom), the sign change indicates \(p\)-type doping, consistent with the higher intrinsic work function of bulk Pt(111) surfaces that typically induces \(p\)-type doping in graphene~\cite{khomyakov2009first, zheng2013interfacial}. This distinction between Pt nanoclusters and extended metallic slabs is largely driven by the under-coordination of Pt in small clusters~\cite{verga2016effect, legesse2017reduced, chan2008first, ramos2013interactions}. Figure~S6 supports these conclusions by displaying charge-density difference plots for the nanoclusters and the Pt(111) slab on graphene. In both nanocluster cases, the isosurfaces show that, in the C–Pt bonding region, electron density accumulates near the graphene layer while depletion occurs at the lower Pt atoms, indicating a net electron transfer from these Pt atoms to graphene. In contrast, the Pt(111) slab exhibits electron depletion closer to graphene and accumulation near the slab, consistent with bulk Pt extracting electrons from graphene and inducing \(p\)-type doping~\cite{khomyakov2009first, zheng2013interfacial}.

Graphene-based chemiresistive gas sensors hinge on variations in electrical resistance that arise from changes in graphene's carrier concentration and/or mobility when the device is exposed to a target gas. Table~\ref{tab:charge_analysis} reveals that hydrogen adsorption profoundly alters the charge‐transfer landscape. For the under-coordinated structures (Pt$_1$, Pt$_9$, and Pt$_{17}$), hydrogen adsorption withdraws electron density from Pt, markedly reducing their \(n\)-type doping of graphene; for a single H atom on the Pt adatom, the net transfer even reverses sign, from \(+5.4\times10^{-3}\,e\) to \(-139\times10^{-3}\,e\). Similarly, adsorption of 6~H atoms atop the Pt$_9$ and Pt$_{17}$ clusters diminishes their charge donation to graphene by approximately 61\;\% and 38\;\%, respectively. At the same time, hydrogen adsorption on Pt renders the adsorption energy to graphene less exothermic, signifying weakened interfacial binding in agreement with MLIP predictions (Figure~\ref{fig:binding_energy}a). In the extended Pt(111) slab, which in its bare form withdraws electron density from graphene, the adsorption of 3 H atoms on the surface donates electron density to Pt, thereby slightly reducing the net charge withdrawal from graphene. However, this reduction amounts to only $6\;\%$, consistent with the Pt–graphene adsorption energy remaining effectively unchanged. This behavior reflects the screening by deeper Pt layers, which attenuate electronic perturbations at the surface after hydrogen adsorption. 

In agreement with MLIP predictions in the previous section, these DFT observations confirm that under-coordinated Pt nanoclusters constitute the most responsive motifs for transducing the H$_2$ adsorption signal to graphene. Hydrogen adsorption on Pt clusters markedly reduces the $n$-type carrier density in graphene—originally supplied by charge donation from the Pt clusters—and concurrently weakens the Pt–graphene interfacial binding, producing a clear electrical transduction signal and potentially increasing the sheet resistance. Experimental studies on Pt‐functionalized graphene H$_2$ sensors report increased device resistance upon H$_2$ exposure~\cite{kim2021tailored}. Although smaller Pt clusters more effectively transduce H$_2$ adsorption into an electrical signal at the Pt–graphene interface, their surface density must remain sufficient to sustain adequate doping levels and signal stability. Increasing Pt loading, however, enhances cluster coalescence and shifts the morphology toward larger aggregates. Therefore, an intermediate Pt loading optimally balances adequate doping with minimal coalescence, preserving the transduction efficiency of under-coordinated Pt clusters while maintaining a robust baseline carrier density, thereby maximizing overall device sensitivity.

\section{Conclusions}
\noindent\textbf{Atomistic‐to‐Device Modeling.} We developed an atomistically resolved, experimentally validated computational framework that bridges \emph{ab initio} quantum accuracy with device-level insight to guide the design of transition metal–functionalized graphene chemiresistive sensors, demonstrating its application to Pt-functionalized graphene for H$_2$ sensing. To this end, we employed an equivariant machine‐learned interatomic potential (MLIP) trained on a diverse C–Pt–H dataset of 5{,}053 structures, achieving energy and force accuracies of $<10\ \mathrm{meV/atom}$ and $<75\ \mathrm{meV/\AA}$, respectively. This enabled nanosecond‐scale molecular dynamics (MD) simulations of Pt crystal growth on $\sim15\times15\ \mathrm{nm}^2$ graphene sheets across 10 coverages (up to 2.5~monolayers). We then simulated the chemiresistive sensing process, evaluated its underlying mechanisms, and developed computational pipelines to extract key kinetic figures of merit for device performance. This data-driven approach enabled the extension of temporal and spatial scales well beyond those accessible through direct DFT calculations.

\noindent\textbf{Crystal Growth.} MLIP-driven MD simulations complemented by transmission electron microscopy (TEM) demonstrated that Pt physical-vapor deposition (PVD) follows a Volmer–Weber growth mode, where adatoms initially nucleate and subsequently coalesce to reduce their surface energy. Domain coalescence was found to proceed via two pathways—filamentary bridge formation and diffusion-driven aggregation—ultimately yielding polycrystalline nanoclusters. Increased Pt loading was observed to reduce the vertical-to-horizontal aspect ratio of island growth, decrease the fraction of under-coordinated edge and corner sites in favor of terrace facets, and also promote structural evolution from disordered configurations to FCC/HCP ordered interiors. These effects were driven by the dominance of Pt–Pt cohesive interactions over local Pt–C interfacial bonding—findings corroborated by TEM and selected area electron diffraction (SAED) analyses. Adsorption energies predicted by the MLIP revealed predominantly non-covalent Pt–graphene interactions. The total adsorption strength increased (with graphene rippling) at higher Pt loadings due to the greater number of C–Pt contacts, despite a reduction in the local binding strength per unit area (i.e., the strength of individual C–Pt bonds). Raman spectroscopy further supported these observations, indicating that increased Pt loading induces greater strain and doping in graphene, while still preserving its structural integrity with minimal lattice disorder or defect formation—an essential condition for maintaining graphene's electronic transport properties in sensing applications.

\noindent\textbf{Gas Sensing Kinetics.} MLIP-driven reactive MD simulations indicated that H$_2$ dissociative chemisorption occurs exclusively on Pt clusters, with negligible H spillover onto pristine graphene. Analysis of both transient and saturation sensing regimes enabled quantification of key gas-sensing metrics, including \emph{response} and \emph{recovery times} as well as \emph{the limit of detection}. Using adsorption and desorption rate constants estimated from MD and applying Langmuir kinetic analysis, we found that lower Pt loadings promote faster response and recovery kinetics. In contrast, an intermediate Pt loading was identified as optimal for minimizing the detection limit, due to a trade-off between the adsorption sticking coefficient and the available reactive surface area.

\noindent\textbf{Electronic Transduction.} Reactive MD simulations further elucidated the sensing mechanism: hydrogen adsorption on Pt modulates Pt–graphene interfacial binding, establishing an indirect electronic transduction pathway for H$_2$ detection at the interface. DFT-based Bader charge analysis revealed that Pt nanoclusters donate electrons to graphene ($n$-type doping), whereas continuous Pt films tend to withdraw electron density ($p$-type doping). Hydrogen adsorption was found to reduce the charge transfer between Pt and graphene, thereby weakening Pt–graphene interactions—consistent with MLIP predictions. This reversible modulation of doping and interfacial adhesion provides a robust mechanism for electrical signal transduction without direct H–graphene chemical bonding. We introduced a related metric, \emph{transduction sensitivity}, to quantify the efficiency of signal transfer from Pt surfaces to the Pt–graphene interface, and found it to be maximal for smaller, dispersed Pt clusters due to shielding effects in larger Pt aggregates. Although smaller clusters transduce H$_2$ adsorption into electrical signals more effectively, their surface density must remain sufficient to maintain stable graphene doping and signal reliability. Consequently, an intermediate Pt loading can optimally balance the need for effective doping while mitigating excessive cluster coalescence, thereby preserving the enhanced transduction efficiency and the faster response and recovery kinetics of smaller Pt clusters, alongside ensuring robust baseline carrier densities.

Overall, the developed machine-learned framework achieves quantum-level accuracy in modeling transition-metal crystal growth on graphene, elucidates the underlying gas sensing mechanisms, and quantitatively correlates key sensing performance metrics with metal loading and morphology. This integrated computational approach offers a predictive pathway that links synthesis conditions directly to sensor performance, thereby enabling the computational optimization of chemiresistive gas sensors.

\section{Methods}
\textbf{DFT Calculations.} DFT calculations were conducted using the generalized gradient approximation (GGA) formulated by Perdew, Burke, and Ernzerhof (PBE) along with projector augmented wave (PAW) pseudo-potentials, using VASP \cite{hohenberg1964inhomogeneous, perdew1996generalized, blochl1994projector, kresse1996efficient}. Collinear spin-polarized calculations were performed, with a plane wave basis cutoff at $520$ eV. Integration over the first Brillouin zone was accomplished using a uniform Monkhorst-Pack mesh with a density of approximately $8.1$ \(\text{Å}^{-1}\) \cite{monkhorst1976special}. The self-consistent field (SCF) loop was terminated when the energy change was $<10^{-5}$ eV. Dispersion interactions were addressed after the VASP calculations using the DFT-D3 method with a cutoff of $6.5$ \AA, consistent with the MLIP cutoff radius. A rational damping function, modified from the original by Becke and Johnson, was employed as implemented in the simple-dftd3 package \cite{ehlert2024simple}. The rational damping function ensures that the dispersion energy approaches a finite value rather than being eliminated at short distances \cite{grimme2011effect}.
\\

\textbf{Machine Learning Interatomic Potential Training.} We employed the Allegro framework for our MLIP \cite{musaelian2023learning, batzner20223}, with a cutoff radius \( r_{\text{max}} = 6.5 \) \AA, maximum order used in spherical harmonics embedding \( l_{\text{max}} = 2 \), full \( O_3 \) parity symmetry, and one tensor product layer with \(32\) features. We utilized a two-body latent multilayer perceptron (MLP) and a later stage latent MLP with hidden dimensions $[32, 64, 128]$ and $[128, 128, 128]$ respectively, both featuring the SiLU (Sigmoid Linear Unit) activation function \cite{elfwing2018sigmoid}. For the final edge energy MLP, a single hidden layer of dimension $32$ without nonlinearity was used. The interatomic distances were embedded using a trainable per-ordered-species-pair radial basis of $6$ Bessel functions and a polynomial cutoff envelope function as specified in \cite{musaelian2023learning}. The employed loss function involved the mean squared error of both atomic forces and per-atom energies, with equal weighting assigned to both components. The learning rate was set to \(4 \times 10^{-4}\), the training batch size \(=5\), and the default Adam optimizer in PyTorch was utilized \cite{kingma2014adam, paszke2019pytorch}. A repulsive Ziegler-Biersack-Littmark (ZBL) term was added to Allegro to enhance the stability of the MLIP at very small interatomic distances \cite{ziegler1985stopping}. 

Our dataset consists of a total of $5,053$ structures: $3,848$ Pt/graphene, $680$ H/Pt/graphene, $464$ Pt-only, and $61$ H-only structures. This resulted in \(121,318\) C, \(54,066\) Pt, and \(18,656\) H local atomic environments. All structures maintain DFT force components \(\leq 40\)  eV/\AA\ and have negative DFT cohesive energies.
\\

\textbf{Minima Hopping.} MH simulations were performed using a modified version of the ASE MH tool, adapted to accommodate Pt$_{\text{N}}$/graphene structures \cite{peterson2014global, larsen2017atomic}. Each MH hop began with a high-temperature ($5000$~K) MD search, with the temperature adaptively rescaled during the run to maintain efficient phase-space exploration. During the MD segment, only the Pt$_{\text{N}}$ cluster was propagated, and the graphene substrate was re‑attached before local energy minimization. The MD segment was terminated once 4 PES minima were traversed. A trial structure was accepted when its energy fell within an adaptively updated threshold that was initialized at $2.5$~eV and tightened (multiplied by~$0.98$) upon acceptance, or relaxed (multiplied by~$1.02$) upon rejection. Accepted structures were then relaxed until all residual forces were $< 0.02$~eV/\AA. This procedure was repeated for $100$ hopping trials, providing extensive sampling of the relevant PES.
\\

\textbf{Molecular Dynamics Simulations.} All MD simulations were performed using LAMMPS with the Allegro pair style \cite{LAMMPS, plimpton1995fast, musaelian2023learning}. The Nosé--Hoover NVT ensemble was employed with a time step of \( dt = 2 \)~fs for Pt crystal growth on graphene, and \( dt = 1 \)~fs for hydrogen sensing simulations. Velocity rescaling was applied every \( 100 \times dt \), and initial velocities were sampled from a Boltzmann distribution corresponding to the target temperature. 

In Pt growth simulations on graphene, the \(x\)–\(y\) coordinates of each new Pt atom were chosen uniformly at random within the supercell, subject to a minimum separation of \(2.8\)~\AA\ from all pre‐existing atoms. For each selected \((x,y)\), the highest local Pt or C atom within an in‐plane cutoff radius of \(5.0\)~\AA\ (to account for graphene corrugations) was identified, and the Pt atom was placed at \(z \ge 3.2\)~\AA\ above that local maximum. This ensured realistic initial adsorption heights while preventing unphysical overlaps. In hydrogen sensing simulations, H$_2$ molecules with an H-H bond length of \(0.75\) \AA\ were initially placed atop the optimized Pt/graphene structure at \(5.0\)--\(20.0\)~\AA\ above the highest Pt atom, while ensuring an initial intermolecular separation of at least \(4.0\) \AA\ between H$_2$ molecules. 

To emulate a rigid substrate, the net linear momentum of the graphene sheet was zeroed at each timestep. In subsequent hydrogen sensing simulations, where H$_2$ molecules were introduced at initial high pressure, the net linear momentum of each Pt cluster was also zeroed at each timestep to mitigate artificial velocity buildup under intense hydrogen impingement, while still allowing drift, local atomic rearrangements, and preserving realistic kinetics.
\\

\textbf{Experimental Pt Functionalization.} Monolayer CVD graphene on SiO$_2$ substrate was commercially obtained from Sigma-Aldrich. All samples were annealed at $200$ °C for $1$ hour under vacuum to remove any residual organics and eliminate adsorbed oxygen and water molecules. This process promotes stronger and more uniform adhesion between graphene and Pt, while suppressing unwanted interfacial reactions. The functionalization of graphene was performed using electron-beam evaporation. Pt deposition was carried out at a rate of \(0.05\)~\AA/s using an Angstrom electron-beam evaporator under a vacuum pressure of \(7 \times 10^{-7}\)~torr. Two different Pt thicknesses ($0.4$ nm and $1.0$ nm) were targeted and measured using a crystal monitor. This ultrathin Pt thickness was optimized to form uniform Pt clusters on the graphene surface without fully coating it, thereby creating distinct catalytic functionalization sites on graphene.

For calibration and characterization, a reference graphene sample on a TEM grid ($2000$ mesh copper grid, Ted Pella Inc.) was placed inside the electron-beam chamber to monitor quality and assist in optimizing Pt deposition.
\\

\textbf{TEM \& Raman Characterization.} The morphology and structural properties of the Pt-functionalized graphene were investigated using TEM. TEM imaging was performed using a JEOL JEM $2100$ microscope with an accelerating voltage of $200$ kV. HRTEM was employed to examine the structural properties of the deposited Pt nanostructures on the graphene surface. A SAED pattern was also obtained to assess the crystallinity of the Pt clusters and determine their crystallographic orientation on graphene. 

Additionally, Raman spectroscopy (Horiba) was performed using a $532$ nm laser to evaluate the quality of graphene before and after Pt functionalization.

\section{Data \& Code Availability}
The DFT training dataset, MLIP configuration, MD simulation videos, and analysis workflows are available at the GitHub repository: \url{https://github.com/UMBC-STEAM-LAB/MLIP-GrPt_CrystGrow-H2_Sensing}.

\section{acknowledgments}
This work was supported by the National Science Foundation through the Division of Materials Research under NSF Grant No. DMR-2213398 and the Department of Energy (DOE) under Grant DE-SC0024236. In addition, the NASA team members were supported by the NASA Polaris program under the Exploration Systems Development Mission Directorate (ESDMD).

\section{REFERENCES}
\bibliography{main}

\section{COMPETING INTERESTS}
The authors declare no competing interests.

\section{ADDITIONAL INFORMATION}
Supplementary information is available for this paper.
\end{document}


\maketitle
\newpage

\begin{figure*}
    \centering
    \includegraphics[width=0.75\textwidth]{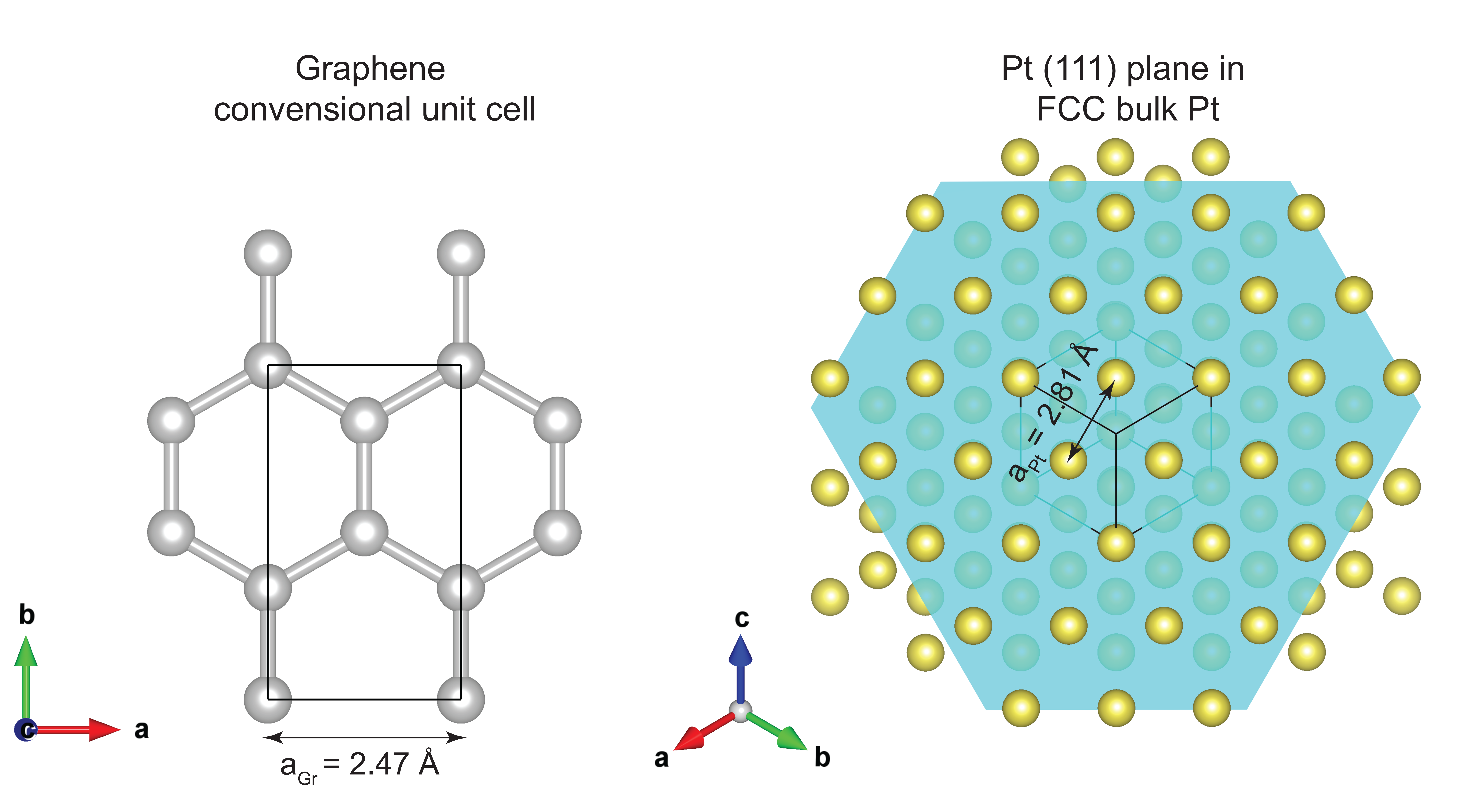}
    \caption{\textbf{Schematic representations of the DFT-relaxed conventional (rectangular) graphene unit cell and the Pt (111) plane in FCC bulk Pt.}}
\label{fig:figS1}
\end{figure*}

Graphene and Pt have lattice mismatch (\(a_{\text{Gr}} = 2.468\,\text{\AA}\) and \(a_{\text{Pt}} = 2.805\,\text{\AA}\)), where \(a_{\text{Gr}}\) represents graphene's lattice constant and \(a_{\text{Pt}}\) represents the nearest neighbor Pt–Pt distance measured in bulk FCC Pt. Assuming a nearly-square graphene supercell of size \(N\) along the zigzag direction, the supercell dimensions along the \(x\) and \(y\) directions (with zigzag along the \(x\)-axis and armchair along the \(y\)-axis) are given by \(N_x = N\) and \(N_y = \frac{N}{\sqrt{3}}\). The area of the graphene conventional unit cell is \(A_{\text{Gr unit}} = a_{\text{Gr}}^2 \sqrt{3}\). Likewise, for a unit cell of a close-packed arrangement of Pt atoms, the area is \(A_{\text{Pt unit}} = a_{\text{Pt}}^2 \sqrt{3}\). Thus, the graphene supercell area is \(A_{\text{Gr super}} = N_x \times N_y \times a_{\text{Gr}}^2 \sqrt{3}\). For a close-packed Pt monolayer covering the graphene area, the number of Pt unit cells is \(A_{\text{Gr super}}/A_{\text{Pt unit}}\), and since each conventional unit cell of the close-packed Pt monolayer contains \(2\) Pt atoms, the total number of Pt atoms is \(N_{\text{Pt}} = A_{\text{Gr super}}/A_{\text{Pt unit}} \times 2\).
Thus, the number of atoms to make a close-packed Pt monolayer that covers the graphene area is \(N_{\text{Pt}} = N_x \times N_y \times \left(\frac{a_{\text{Gr}}}{a_{\text{Pt}}}\right)^2 \times 2\). For example, if \(N_x = 60\) and \(N_y = 35\), then \(N_{\text{Pt}} = 60 \times 35 \times \left(\frac{2.468\,\text{\AA}}{2.805\,\text{\AA}}\right)^2 \times 2 \approx 3249\).


\begin{figure*}
    \centering
    \includegraphics[width=0.94\textwidth]{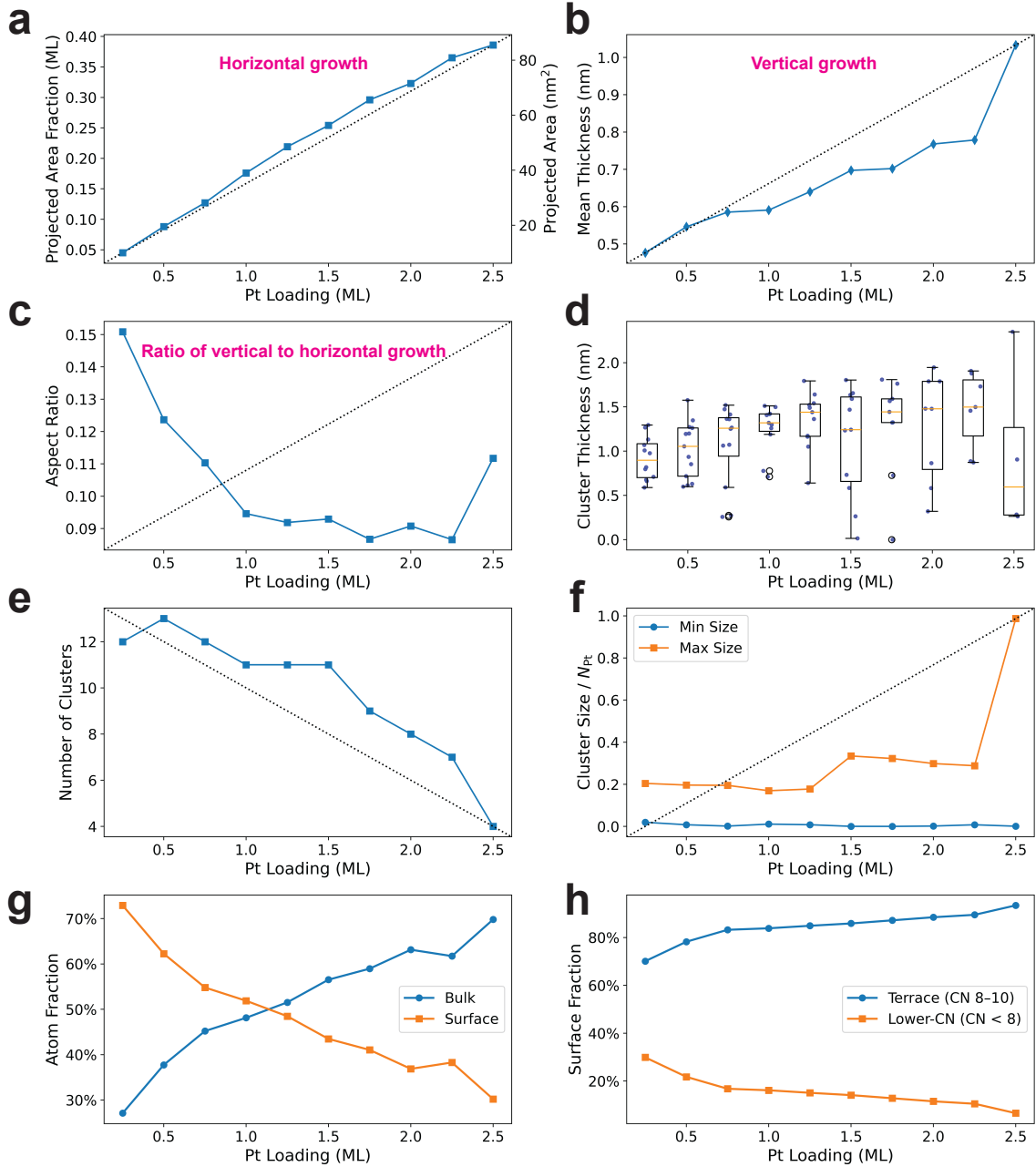}
    \caption{\textbf{Cluster Properties versus Pt Loading.} (\textbf{a}) Projected area $A_{proj}$ in nm$^2$ and as a fraction of the underlying graphene area. (\textbf{b}) Mean thickness of Pt domains, $\bar t = \frac{1}{N_{\mathrm{Pt}}} \sum_{k=1}^{N_{\mathrm{clusters}}} \sum_{i=1}^{N_{\mathrm{Pt},k}} (z_{i,k} - z_{\min,k})$. (\textbf{c}) Aspect ratio $\bar t/\sqrt{A_{proj}}$. (\textbf{d}) Boxplots of Pt‐cluster thickness versus Pt loading: the central line is the median, the box spans the 25th–75th percentiles (IQR), whiskers extend to $1.5 \times$ IQR (caps at their ends), and blue jittered dots are the data points. (\textbf{e}) Number of formed distinct Pt clusters. (\textbf{f}) Sizes of minimum and maximum clusters normalized by $N_{\mathrm{Pt}}$. (\textbf{g}) Bulk vs.\ surface atom fractions as a function of Pt loading; surface atoms are those with coordination number \(\mathrm{CN}<11\), bulk those with \(\mathrm{CN}=11\text{–}12\). (\textbf{h}) Breakdown of surface sites into terrace atoms (\(\mathrm{CN}=8\text{–}10\)) and lower-coordination atoms (\(\mathrm{CN}<8\)), plotted as their fraction of all surface atoms versus Pt loading.}
    \label{fig:figS2}
\end{figure*}

\newpage

\begin{figure}
    \centering
    \includegraphics[width=0.49\textwidth]{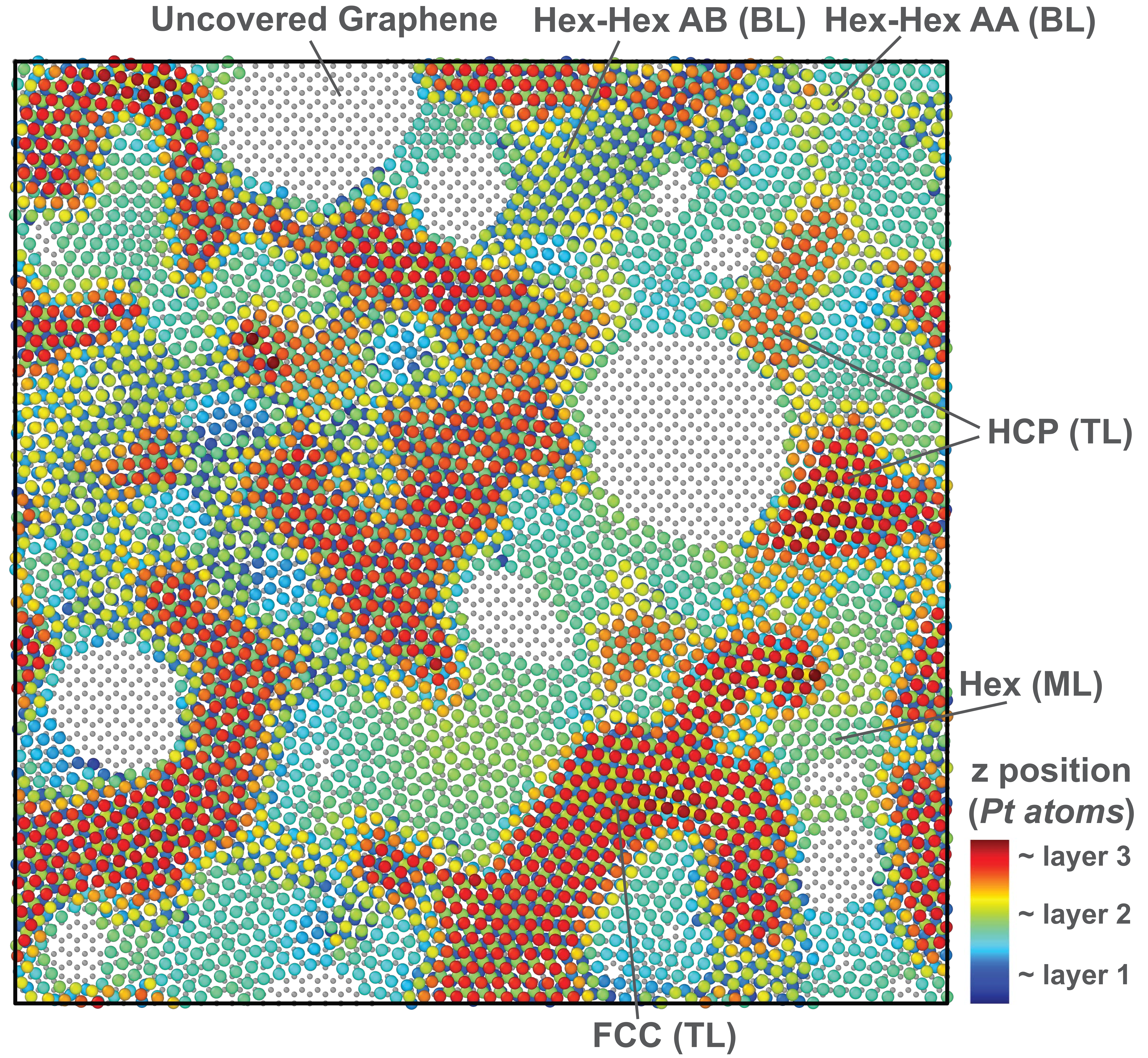}
    \caption{\textbf{MLIP-Optimized 2D Pt Domains on Graphene.} Final configuration from MD annealing following a single-batch deposition of 2.00 ML Pt on graphene. The color map indicates the $z$-coordinate of the Pt atoms.}
    \label{fig:figS3}
\end{figure}

\begin{table}
  \centering
    \caption{Key structural and energetic properties of Pt on graphene at a $2.00$~ML loading, comparing 3D nanoclusters and 2D film morphologies.}
  \label{tab:comparison}
  \begin{tabular*}{0.94\textwidth}{@{\extracolsep{\fill}}lcc}
    \hline
    \textbf{Pt Morphology}   & \textbf{3D Nanocluster} & \textbf{2D Film} \\
    \hline
    $A_{\text{surf}}$ (ML)                & 1.17 & 1.95 \\
    $A_{\text{proj}}$ (ML)              & 0.32 & 0.93 \\
    $A_{\text{expos}}$ (ML)              & 0.85 & 1.02 \\
    $E^{\text{form}}_{\text{Pt}}$ (eV/Pt)           & -5.49 & -5.27 \\
    $E^{\text{surf}}_{\text{Pt}}$ (keV)          & 3.10 & 4.88 \\
    $E^{\text{ads}}_{\text{Pt}}$ (keV)          & -0.32 & -0.63 \\
    $E^{\text{ads}}_{\text{Pt}} / A_{\text{proj}}$ (eV/nm$^2$)      & -4.40 & -3.05 \\
    \hline
  \end{tabular*}
\end{table}

\newpage

\begin{table*}
  \centering
  \caption{DFT+D3 cohesive energies and lattice parameters of various Pt bilayer structural models.}
  \label{tab:pt_models}
  \begin{tabular*}{0.94\textwidth}{@{\extracolsep{\fill}}lcccc}
    \hline
    \textbf{Model} & \textbf{Cohesive Energy (eV/atom)} & \textbf{$a$ (\AA)} & \textbf{$b$ (\AA)} & \textbf{$d$ (\AA)} \\
    \hline
    sq-sq-top        & -4.858 & 2.60 & 2.60 & 2.53 \\
    sq-sq-center     & -4.954 & 2.64 & 2.64 & 2.08 \\
    hex-hex-bridge   & -5.186 & 2.68 & 2.33  & 2.44 \\
    hex-hex-top      & -5.188 & 2.68 & 2.32  & 2.63 \\
    hex-hex-center   & -5.193 & 2.69 & 2.33  & 2.41 \\
    \hline
  \end{tabular*}
\end{table*}

\begin{figure*}
    \centering
    \includegraphics[width=0.94\textwidth]{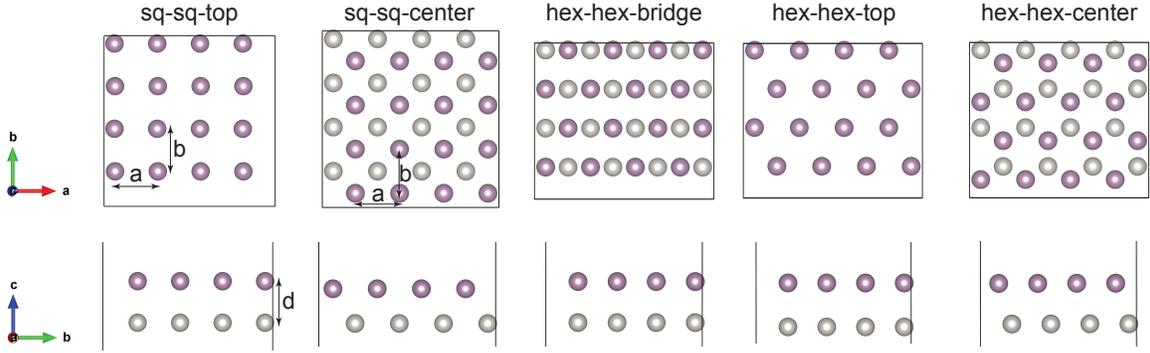}
    \caption{\textbf{Schematic representations of DFT-relaxed, free-standing Pt bilayer models.} The figure displays Pt bilayer models with both square (sq) and hexagonal (hex) in-plane arrangements. Distinct stable second-layer stacking configurations (top, center, and bridge) are highlighted in violet atop the silver-colored first layer. Cohesive energies and lattice constants are detailed in Table S2.}
\label{fig:figS4}
\end{figure*}

\begin{figure*}
    \centering
    \includegraphics[width=0.94\textwidth]{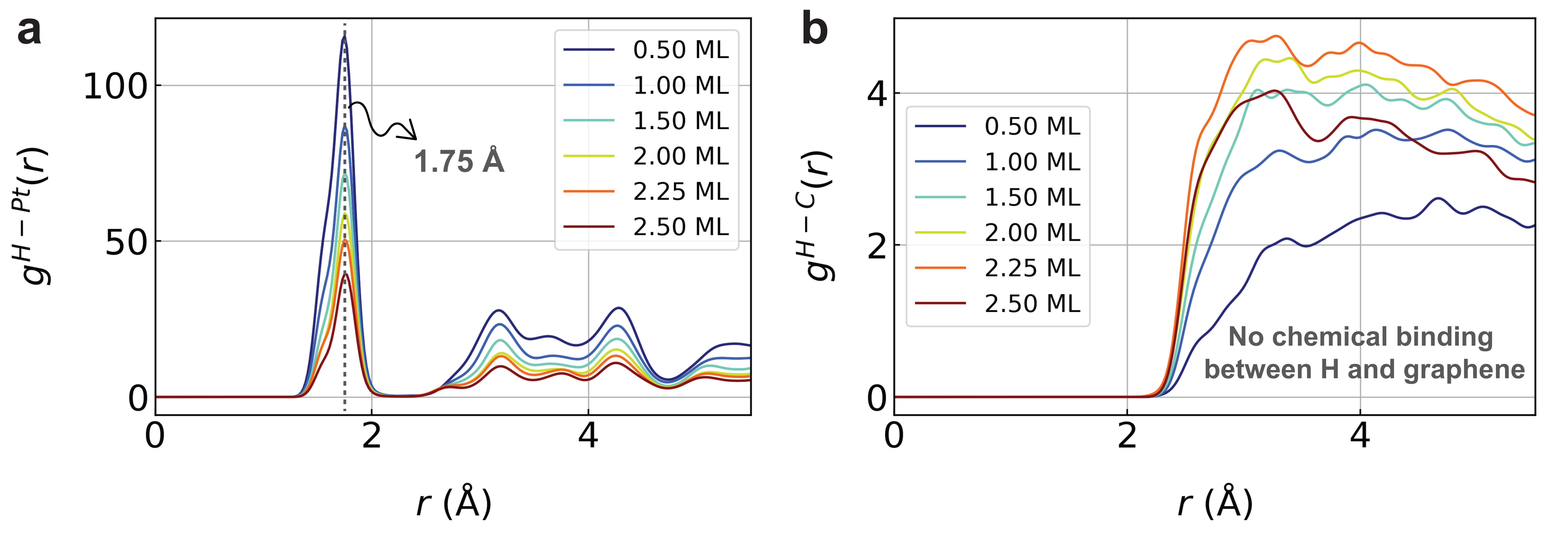}
    \caption{\textbf{H-Pt and H-C radial distribution functions (RDFs).} \textbf{a} H–Pt RDF and \textbf{b} H–C RDF at various Pt loadings (obtained after relaxing the final structures from MD).}
    \label{fig:figS5}
\end{figure*}

\begin{figure*}
    \centering
    \includegraphics[width=0.94\textwidth]{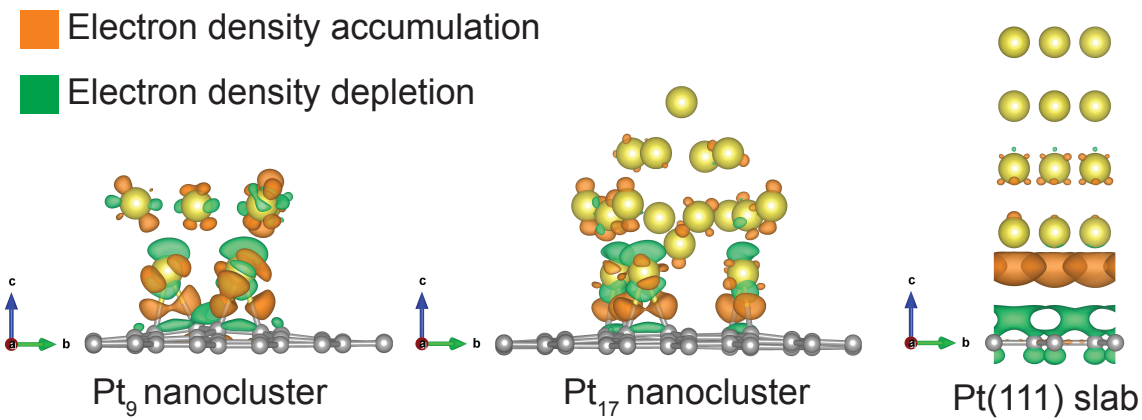}
    \caption{\textbf{Charge Density Difference.} The figure shows charge density differences for 3 Pt/graphene configurations: Pt\(_9\) and Pt\(_{17}\) nanoclusters, and a 4-layer Pt(111) slab. Orange isosurfaces indicate electron accumulation, while green denotes electron depletion. Isosurface values are \(0.005\) e/\AA\(^3\) for the nanoclusters and \(0.0003\) e/\AA\(^3\) for the slab.}
    \label{fig:figS6}
\end{figure*}

\newpage

\section*{Derivation of the transient-period analytical solution}

We begin with the total‐atom balance and ideal‐gas relation:
\begin{align}
&2\,N^{(g)}_{\mathrm{H}_2}(t) + N_H^{(s)}(t) = 2N^{(g)}_{\mathrm{H}_2}(0),
\quad\Longrightarrow\quad
N^{(g)}_{\mathrm{H}_2}(t) = \frac{2N^{(g)}_{\mathrm{H}_2}(0) - N_H^{(s)}(t)}{2}, 
\label{eq:mass_balance}\\
&P(t) \;=\;\frac{k_{\mathrm{B}}T}{V}\,N^{(g)}_{\mathrm{H}_2}(t)
\;=\;\frac{k_{\mathrm{B}}T}{V}\,\frac{2N^{(g)}_{\mathrm{H}_2}(0) - N_H^{(s)}(t)}{2}
\;=\;P_0 \;-\;\frac{k_{\mathrm{B}}T}{2V}\,N_H^{(s)}(t),
\label{eq:pressure_general}
\end{align}
where we define the initial pressure
\[
P_0 \;=\;\frac{N^{(g)}_{\mathrm{H}_2}(0)k_{\mathrm{B}}T}{V}.
\]
Introducing the fractional coverage 
\(
\theta(t)=N_H^{(s)}(t)/N_{\mathrm{sites}}
\)
and the dimensionless coupling
\(
b = \tfrac{N_{\mathrm{sites}}}{2N^{(g)}_{\mathrm{H}_2}(0)},
\)
we rewrite the pressure as
\begin{equation}
P(t) \;=\; P_0 \Bigl[1 - b\,\theta(t)\Bigr].
\label{eq:pressure_theta}
\end{equation}

Under Langmuir kinetics with second‐order adsorption and recombinative desorption,
\begin{equation}
\frac{d\theta}{dt}
= k_{\mathrm{ads}}\,P(t)\,\bigl(1-\theta\bigr)^2
- k_{\mathrm{des}}\,\theta^2.
\label{eq:theta_full}
\end{equation}
Substitute \(P(t)\) from \eqref{eq:pressure_theta}:
\[
\frac{d\theta}{dt}
= k_{\mathrm{ads}}\,P_0\,(1-b\theta)\,(1-\theta)^2
- k_{\mathrm{des}}\,\theta^2.
\]
\medskip

\noindent\textbf{1. Low‐coverage expansion.}\\
For \(\theta\ll1\), expand each factor to \(O(\theta)\):
\begin{align*}
(1-\theta)^2 &= 1 - 2\theta + O(\theta^2),\\
1 - b\,\theta   &= 1 - b\,\theta,\\
k_{\mathrm{des}}\,\theta^2 &= O(\theta^2)\;\to\;0.
\end{align*}
Thus, retaining only linear terms,
\[
\frac{d\theta}{dt}
\approx k_{\mathrm{ads}}\,P_0\,(1 - b\,\theta)\,(1 - 2\theta)
= k_{\mathrm{ads}}P_0
\Bigl[1 - (b+2)\,\theta\Bigr].
\]
Rearrange into standard first‐order form:
\begin{equation}
\frac{d\theta}{dt} + k_{\mathrm{eff}}\,\theta = k_{\mathrm{ads}}P_0,
\qquad
k_{\mathrm{eff}} = (b+2)\,k_{\mathrm{ads}}P_0.
\label{eq:linear_ode}
\end{equation}

\noindent\textbf{2. Solution via integrating factor.}\\
The integrating factor is \(\exp(k_{\mathrm{eff}} t)\). Multiplying \eqref{eq:linear_ode} by this factor,
\[
\frac{d}{dt}\Bigl[e^{k_{\mathrm{eff}}t}\,\theta\Bigr]
= k_{\mathrm{ads}}P_0\,e^{k_{\mathrm{eff}}t}.
\]
Integrate from \(0\) to \(t\) with \(\theta(0)=0\):
\[
e^{k_{\mathrm{eff}}t}\,\theta(t)
= \frac{k_{\mathrm{ads}}P_0}{k_{\mathrm{eff}}}
\Bigl[e^{k_{\mathrm{eff}}t}-1\Bigr],
\]
hence
\begin{equation}
\theta(t)
= \frac{k_{\mathrm{ads}}P_0}{k_{\mathrm{eff}}}\Bigl[1-e^{-k_{\mathrm{eff}}t}\Bigr]
= \frac{1}{b+2}\Bigl[1-e^{-k_{\mathrm{eff}}t}\Bigr].
\label{eq:theta_solution}
\end{equation}

\noindent\textbf{3. Recovering \(N_H^{(s)}(t)\) and \(P(t)\).}\\
Using \(N_H^{(s)}=N_{\mathrm{sites}}\,\theta\) and \eqref{eq:pressure_theta}:
\begin{align}
N_H^{(s)}(t)
&= \frac{N_{\mathrm{sites}}}{b+2}\,\bigl[1 - e^{-k_{\mathrm{eff}}\,t}\bigr]
\label{eq:Ns_exp}\\
P(t)
&= P_0\Bigl[1 - \frac{b}{b+2}\bigl(1 - e^{-k_{\mathrm{eff}}\,t}\bigr)\Bigr]
\label{eq:P_exp}
\end{align}
where
\[
b = \frac{N_{\mathrm{sites}}}{2N^{(g)}_{\mathrm{H}_2}(0)},
\quad
k_{\mathrm{eff}} = (b+2)\,k_{\mathrm{ads}}\,P_0.
\]

Equations \eqref{eq:pressure_theta} furnish the approximate exponential kinetics observed for the transient response.

\section*{Extraction of \(k_{\mathrm{eff}}\) via log‐linear fit}

Starting from Eq.~\eqref{eq:Ns_exp}, define the normalized residual
\begin{equation}
R(t) \;=\; 1 \;-\; \frac{b+2}{N_{\mathrm{sites}}}\,N_H^{(s)}(t)
\;=\; e^{-k_{\mathrm{eff}}\,t}\,.
\label{eq:R_def}
\end{equation}
Taking the natural logarithm gives
\begin{align}
\ln R(t)
&= \ln\!\Bigl[\,1 - \tfrac{b+2}{N_{\mathrm{sites}}}\,N_H^{(s)}(t)\Bigr]
\;=\; -\,k_{\mathrm{eff}}\,t\,.
\label{eq:lnR_linear}
\end{align}
Thus, plotting \(\ln R(t)\) versus \(t\) yields a straight line:
\begin{equation}
\ln R(t) = m\,t + c,
\label{eq:lnR_fit}
\end{equation}
with slope
\[
m = -\,k_{\mathrm{eff}}
\quad\Longrightarrow\quad
k_{\mathrm{eff}} = -\,m.
\] 

\newpage

\begin{table*}
  \centering
    \caption{SAED d-spacings and their corresponding reflection planes.}
  \label{tab:pt_models}
  \begin{tabular*}{0.94\textwidth}{@{\extracolsep{\fill}}lcc}
    \hline
    \textbf{Ring No.} & \textbf{d-spacing (nm)} & \textbf{(hkl)} \\
    \hline
    1   & 0.31 &  FCC (110) \\
    2   & 0.25 & SC (100)\\
    3   & 0.21 &  Gr (100)\\
    4   & 0.15 &  SC (110)\\
    5   & 0.13 &  FCC (220)\\
    6   & 0.12 & Gr (110)\\
    7   & 0.108 & Gr (200)\\
    \hline
  \end{tabular*}
\end{table*}

\begin{figure*}
    \centering
    \includegraphics[scale=0.48]{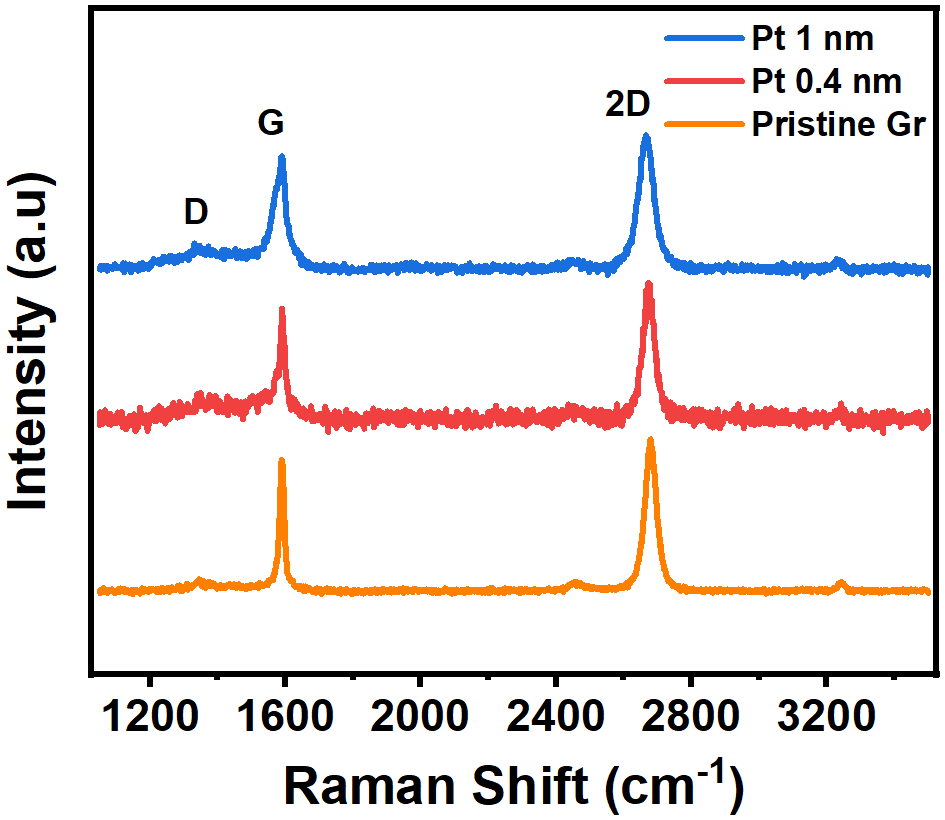}
    \caption{Raman shift for pristine graphene (orange), Pt 0.4 nm (red), and Pt 1 nm (blue).}
    \label{fig:Raman_Pt.png}
\end{figure*}
